\documentclass{article}
\usepackage[hyphens]{url}
\usepackage[utf8]{inputenc}
\usepackage[sort&compress, numbers]{natbib}
\usepackage{graphicx}
\usepackage{authblk}
\usepackage[english]{babel}
\usepackage{setspace}
\usepackage{fancyhdr}
\usepackage{array}
\usepackage[margin=1in]{geometry}
\usepackage{float}
\usepackage{datetime}
\usepackage{amsmath}
\usepackage{amssymb}
\usepackage{bbm}
\usepackage{bm}
\usepackage[final]{pdfpages}
\usepackage{longtable}
\usepackage{chngcntr}
\usepackage{placeins}
\usepackage{microtype}
\usepackage{tikz}
\usepackage{makecell}
\usepackage{hyperref}
\usepackage{subcaption}
\usepackage{rotating}
\usepackage{soul}
\usepackage{tabularx}

\hypersetup{
    colorlinks = true,
    citecolor = {blue},
    urlcolor = {blue},
    menucolor = {blue},
    linkcolor = {blue}
}

\makeatletter
\def\@author{}
\renewcommand\@author{\ifx\AB@affillist\AB@empty\AB@author\else
      \ifnum\value{affil}>\value{Maxaffil}\def\rlap##1{##1}%
    \AB@authlist \\*[0.1cm] \small on behalf of EU-PEARL (EU Patient-cEntric clinicAl tRial pLatforms) Consortium \\[\affilsep]\AB@affillist
    \else  \AB@authors\fi\fi}
\makeatother

\title{Decision rules for identifying combination therapies in open-entry, randomized controlled platform trials}

\author[1]{Elias Laurin Meyer}
\author[2]{Peter Mesenbrink}
\author[3]{Cornelia Dunger-Baldauf}
\author[3,4]{Ekkehard Glimm}
\author[2]{Yuhan Li}
\author[1,*]{Franz König}
\affil[1]{Center for Medical Statistics, Informatics, and Intelligent Systems, Medical University of Vienna, Austria}
\affil[2]{Novartis Pharmaceuticals Corporation, One Health Plaza, East Hanover, NJ, USA}
\affil[3]{Novartis Pharma AG, Basel, Switzerland}
\affil[4]{Institute of Biometry and Medical Informatics, University of Magdeburg, Germany}
\affil[*]{Correspondence: franz.koenig@meduniwien.ac.at; Tel.: +43-1-40400-74800}
\date{}         
\begin{document}

\maketitle

\begin{abstract}

Platform trials have become increasingly popular for drug development programs, attracting interest from statisticians, clinicians and regulatory agencies. Many statistical questions related to designing platform trials - such as the impact of decision rules, sharing of information across cohorts, and allocation ratios on operating characteristics and error rates - remain unanswered. In many platform trials, the definition of error rates is not straightforward as classical error rate concepts are not applicable. For an open-entry, exploratory platform trial design comparing combination therapies to the respective monotherapies and standard-of-care, we define a set of error rates and operating characteristics and then use these to compare a set of design parameters under a range of simulation assumptions. When setting up the simulations, we aimed for realistic trial trajectories, such that e.g. a priori we do not know the exact number of treatments that will be included over time in a specific simulation run as this follows a stochastic mechanism. Our results indicate that the method of data sharing, exact specification of decision rules and a priori assumptions regarding the treatment efficacy all strongly contribute to the operating characteristics of the platform trial. Furthermore, different operating characteristics might be of importance to different stakeholders. Together with the potential flexibility and complexity of a platform trial, which also impact the achieved operating characteristics via e.g. the degree of efficiency of data sharing, this implies that utmost care needs to be given to evaluation of different assumptions and design parameters at the design stage. \newline
\newline

\end{abstract}

\section{Introduction} \label{Introduction}

The goal to test as many investigational treatments as possible over the shortest duration, which is influenced by both recent advances in drug discovery and biotechnology and the ongoing global pandemic due to the SARS-CoV-2 virus \citep{Kunz2020, Stallard2020Covid, Dodd2020, Horby2020, Angus2021, Macleod2021}, has made master protocols and especially platform trials an increasingly feasible alternative solution to the time-consuming sequence of classical randomized controlled trials \citep{Woodcock2017, Angus2019, Park2019, Park2020, Meyer2020, Israel2021}. Platform trial designs allow for the evaluation of one or more investigational treatments in the study population(s) of interest within the same clinical trial, as compared to traditional randomized controlled trials, which usually evaluate only one investigational treatment in one study population. When cohorts share key inclusion/exclusion criteria, trial data can easily be shared across such sub-studies. In practice, setting up a platform trial potentially may require additional time due to operational and statistical challenges. However, simulations have shown that platform trials can be superior to classical trial designs with respect to various operating characteristics which include the overall study duration. In a setting where only few new agents are expected to be superior to standard of care, \citet{Saville2016} investigated the operating characteristics of adaptive Bayesian platform trials using binary endpoints compared with a sequence of ``traditional" trials, i.e. trials testing only one hypothesis, and found that platform trials perform dramatically better in terms of the number of patients and time required until the first superior investigational treatment has been identified. Using real data from the 2013-2016 Ebola virus disease epidemic in West Africa, \citet{Brueckner2018} investigated the operating characteristics of various multi-arm multi-stage and two-arm single stage designs, as well as group-sequential two-arm designs, and found that designs with frequent interim analyses outperformed single-stage designs with respect to average trial duration and sample size when fixing the type 1 error and power. When having a pool of investigational agents available, which should all be tested against a common shared control, and using progression-free survival as the efficacy endpoint, \citet{Yuan2016} found that average trial duration and average sample size are drastically reduced when using a multi-arm, Bayesian adaptive platform trial design using response-adaptive randomization compared with traditional two-arm trials evaluating one agent at a time. \citet{Hobbs2018_2} reached a similar conclusion when comparing a platform trial with a binary endpoint and futility monitoring based on Bayesian predictive probabilities with a sequence of two-arm trials. \citet{Tang2018} investigated a phase II setting in which several monotherapies are combined with several backbone therapies and tested in a single-arm manner. Assuming different treatment combination effects, they found that their proposed Bayesian platform design with adaptive shrinkage has a lower average sample size in the majority of scenarios investigated and always a higher percentage of correct combination selections when compared with a fully Bayesian hierarchical model and a sequence of Simon's two-stage designs. \citet{Ventz2017} proposed a frequentist adaptive platform (so called "rolling-arms design") design as an alternative to sequences of two-arm designs and Bayesian adaptive platform designs, which is much simpler than the Bayesian adaptive platform designs in that it uses equal allocation ratios and simpler and established decision rules based on group sequential analysis. The authors found that performance under different treatment effect assumptions and a set of general assumptions was comparable to, if not slightly better than Bayesian adaptive platform designs and much better than a sequence of traditional two-arm designs in terms of average sample size and study duration. For a comprehensive review on the evolution of master protocol clinical trials and the differences between basket, umbrella and platform trials, see \citet{Meyer2020}. In this article, we explore the impact of both decision rules and assumptions on the nature of the treatment effects and availability of treatments on certain operating characteristics of an open-entry, cohort platform trial with multiple study arms. Such a platform design makes sense in any context in which multiple pair-wise comparisons are necessary to advance a compound, whereby some comparisons are based on treatments unique to different cohorts and some comparisons are based on treatments common to all cohorts (and any mixture thereof). An example for such a context would be a drug development program in Nonalcoholic steatohepatitis (NASH) trying to advance two-compound combination therapies with a common backbone therapy. \newline
\newline
The article is organized as follows: In Section \ref{Methods}, we describe the trial design under investigation, the different testing strategies as well as the investigated operating characteristics. In Section \ref{Simulations}, we discuss the simulation setup and present and discuss the results of the different simulation scenarios. We conclude with a general discussion in Section \ref{Discussion}.


\section{Methods} \label{Methods}

\subsection{Platform Design} \label{Trial Design}

\begin{sidewaysfigure}[ht]
\includegraphics[scale=0.65]{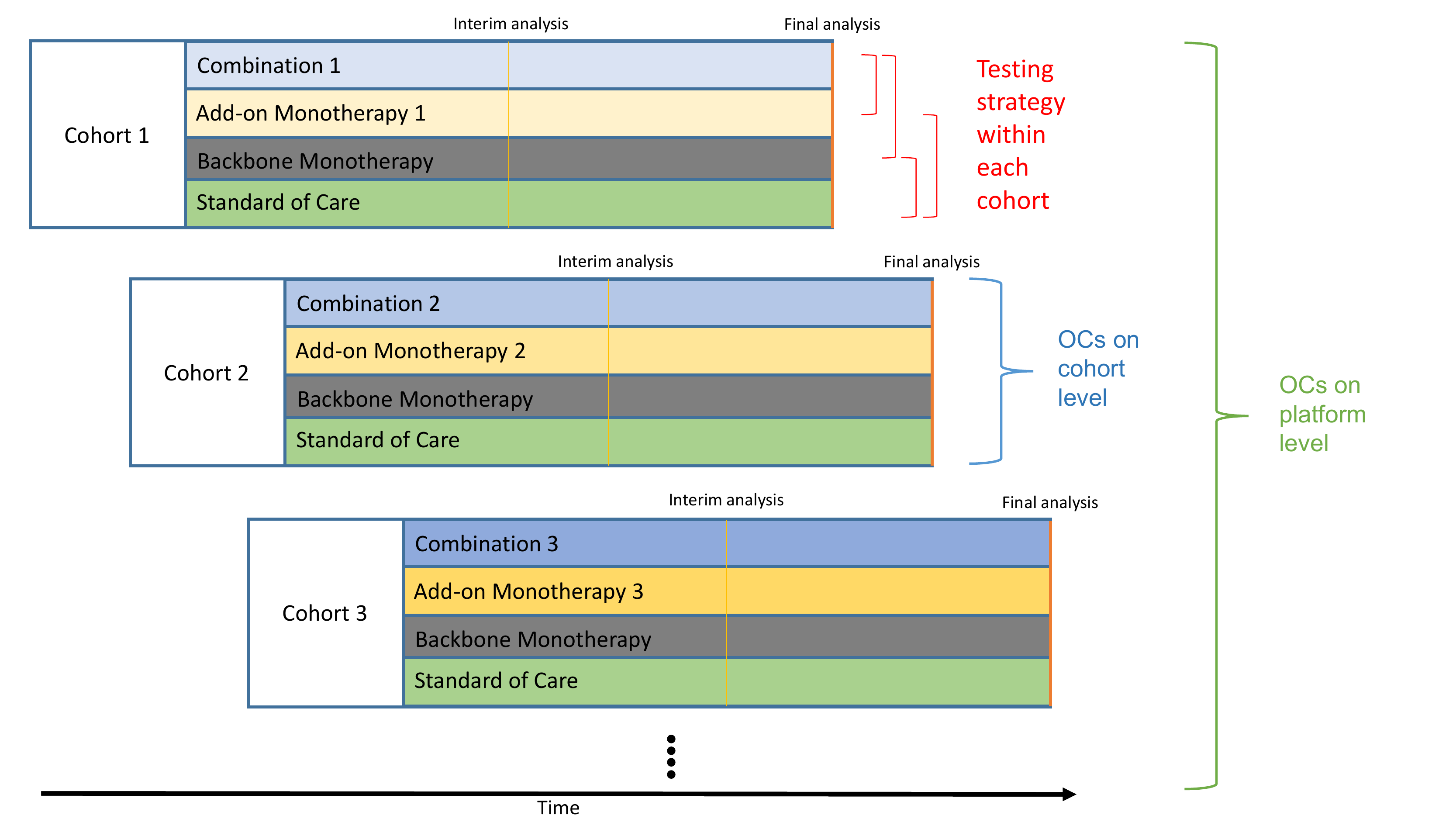}
\caption{Schematic overview of the proposed platform trial design. New cohorts consisting of a combination therapy arm, a monotherapy arm using the same compound in every cohort (referred to as "backbone monotherapy"), an add-on monotherapy arm which is different in every cohort and a SoC arm are entering the platform over time. While the add-on monotherapy and therefore the combination therapy is different in every cohort (as indicated by the differently shaded colors), the backbone monotherapy and SoC are the same in every cohort (as indicated by the same colors). Each cohort has an interim analysis after about half of the initially planned sample size, after which the cohort can be stopped for early efficacy or futility. The red brackets indicate the testing strategy within each cohort, i.e. comparison of combination therapy against both monotherapies and both monotherapies against SoC. We differentiate between per-cohort and per-platform operating characteristics (OCs).}
\label{fig:trialdesign}
\end{sidewaysfigure}

We investigated an open-entry, exploratory cohort platform study design with a binary endpoint evaluating the efficacy of a two-compound combination therapy compared to the respective monotherapies and the standard-of-care (SoC). After an initial inclusion of one cohort, we allow new cohorts to enter the platform trial over time until a maximum number of cohorts is reached. We assume different cohorts share common key inclusion/exclusion criteria and patients are always drawn from the same population, such that all eligible patients can be randomized to any cohort. Each cohort consists of four arms: combination therapy, monotherapy A, monotherapy B and SoC. Monotherapy A is the same for all cohorts (further referred to as "backbone monotherapy"), while monotherapy B (further referred to as "add-on monotherapy X") is different in every cohort X. The combination of monotherapies A and B is called {\it combination therapy}. See Figure \ref{fig:trialdesign} for a schematic overview of the proposed trial design. To show that the backbone and SoC treatments do not change during the platform trial, we use the same color coding of grey and green for all cohorts, respectively. The x-axis shows the calendar time. At any point in time, new cohorts could enter the platform trial. In contrast to a simulation of a classical RCT, here different response rates for the same treatment arms in different cohorts can be assumed. The assignment of response rates to treatment arms is described in more detail in appendix \ref{Efficacy}. We conduct one interim analysis for every cohort (indicated by the vertical yellow line in Figure \ref{fig:trialdesign}), on the basis of which the cohort might be stopped early for either futility or efficacy. The platform trial ends if there is no active, recruiting cohort left. The proposed trial design is suggested for phase II trials. Subsequent phase III studies would follow if a combination therapy is graduated successfully. Following a rigorous interpretation of the current FDA and EMA regulatory guidelines \citep{FDA_Codevelopment, EMA_Codevelopment}, superiority of the combination therapy over both monotherapies and superiority of both monotherapies over SoC needs to be demonstrated. Therefore, we test the combination therapy against both of the monotherapies and both of the monotherapies against SoC, resulting in four comparisons (indicated by the red "testing strategy" lines in Figure \ref{fig:trialdesign}). 


\begin{sidewaysfigure}[ht]
\includegraphics[scale=0.65]{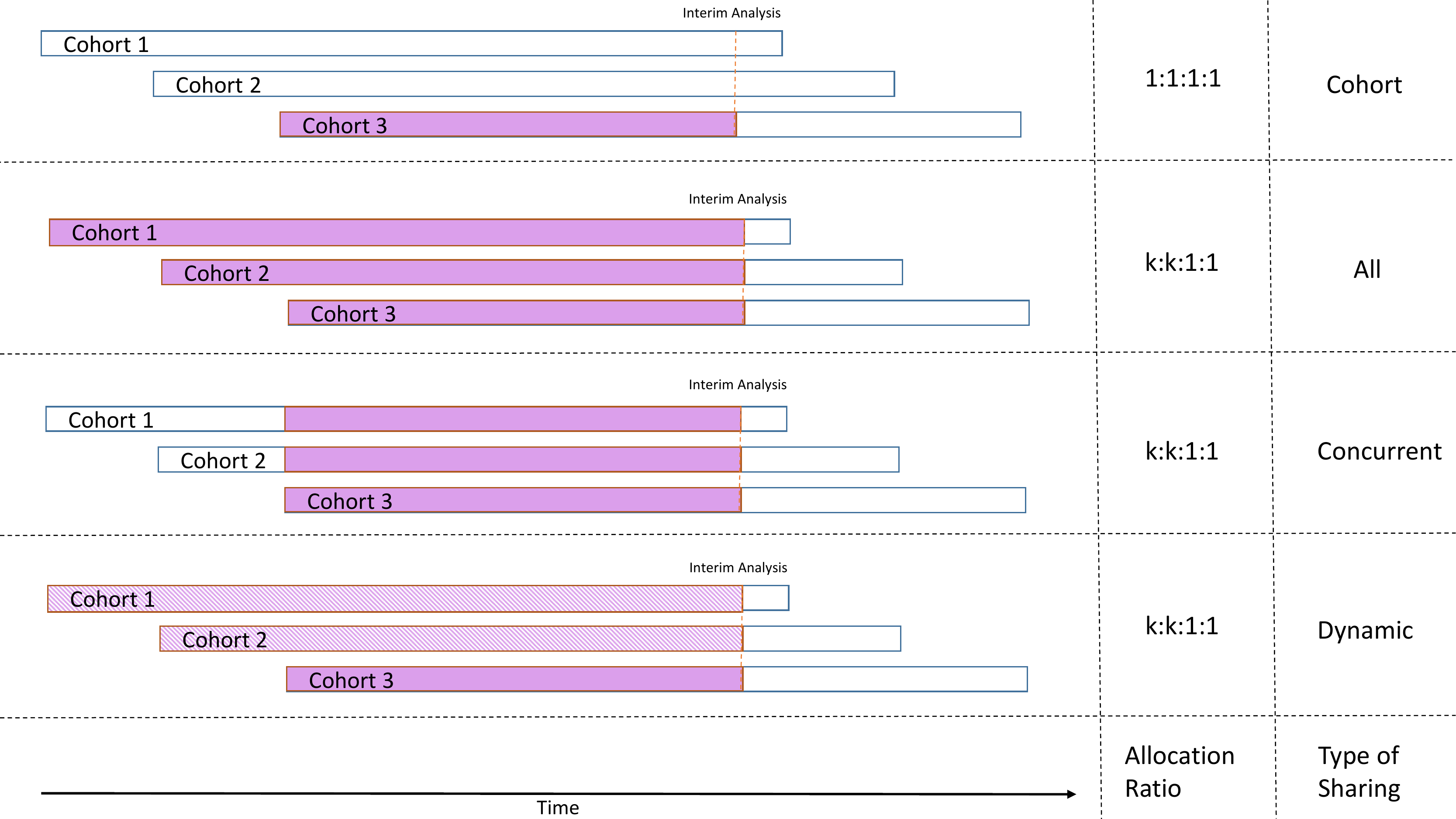}
\caption{Schematic overview of the different levels of sharing. No sharing happens if only "cohort" data are used. If sharing "all" data, whenever in any cohort an interim or final analysis is performed, all SoC and backbone monotherapy data available from all cohorts are used. If sharing only "concurrent" data, whenever in any cohort an interim or final analysis is performed, all SoC and backbone monotherapy data that was collected during the active enrollment time of the cohort under investigation are used. If sharing "dynamically", whenever in any cohort an interim or final analysis is performed, the degree of data sharing of SoC and backbone monotherapy data from other cohorts increases with the homogeneity of the observed response rate of the respective arms. A solid fill represents using data 1-to-1, while a dashed fill represents using discounted data (for more information see appendix \ref{dynamic_borrowing}). If at any given time there are $k$ active cohorts, the allocation ratio is $1:1:1:1$ in case of no data sharing and $k:k:1:1$ otherwise (combination : add-on monotherapy : backbone monotherapy : SoC). This allocation ratio is updated for all active cohorts every time the number of active cohorts $k$ changes due to dropping or adding a new cohort.}
\label{fig:sharing}
\end{sidewaysfigure}


\subsection{Bayesian Decision Rules} \label{Decision Rules}

As we discussed in section \ref{Trial Design}, for testing the efficacy of the combination therapy, four pair-wise comparisons are performed. In this article, for every one of the comparisons, we consider Bayesian decision rules based on the posterior distributions of the response rates of the respective study arms \citep{Jiang2020}. In principle, any other Bayesian (e.g. predictive probabilities, hierarchical models, etc.) or frequentist rules (based on p-values, point estimates or confidence intervals) could be used. While these decision rules are based on fundamentally different paradigms, they might translate into the exact same stopping rules, e.g. with respect to the observed response rate \citep{Gallo2014, Stallard2020}. Decision rules based on posterior distributions used vague, independent Beta(1/2, 1/2) priors for all simulation results presented in this paper; however, please note that independence of the vague priors is a strong assumption and its violation can lead to selection bias as pointed out by many authors \citep{Dawid1994, Senn2008, Mandel2009}.We differentiate two types of decisions for cohorts: ``GO" (graduate combination therapy, i.e. declare combination therapy successful) and ``STOP" (stop evaluation of combination therapy and do not graduate, i.e. declare the combination therapy unsuccessful). Generally, we consider decision rules of the following form:

\begin{align}
\begin{split}
\label{eq:DR}
\text{GO, if } & (P(\pi_{C} > \pi_{A} + \delta_{CA}^{E,T} | Data) > \gamma_{CA}^{E,T}) \ \wedge \\
               & (P(\pi_{C} > \pi_{B} + \delta_{CB}^{E,T} | Data) > \gamma_{CB}^{E,T}) \ \wedge \\
               & (P(\pi_{A} > \pi_{S} +  \delta_{AS}^{E,T} | Data) > \gamma_{AS}^{E,T}) \ \wedge \\
               & (P(\pi_{B} > \pi_{S} +  \delta_{BS}^{E,T} | Data) > \gamma_{BS}^{E,T}) \\
               \\
\text{STOP, if } & (P(\pi_{C} > \pi_{A} + \delta_{CA}^{F,T} | Data) < \gamma_{CA}^{F,T}) \ \vee \\
                 & (P(\pi_{C} > \pi_{B} + \delta_{CB}^{F,T} | Data) < \gamma_{CB}^{F,T}) \ \vee \\
                 & (P(\pi_{A} > \pi_{S}  + \delta_{AS}^{F,T} | Data) < \gamma_{AS}^{F,T}) \ \vee \\
                 & (P(\pi_{B} > \pi_{S}  + \delta_{BS}^{F,T} | Data) < \gamma_{BS}^{F,T}) \\
                 \\
\end{split}
\end{align} 

whereby $\pi_x$ denotes the response rate in treatment arm $x$ ($x \in \{C, A, B, S\}$) and $T \in {1,2}$ denotes the analysis time point. At interim ($T = 1$), if neither a decision for early efficacy or futility is made, the cohort continues unchanged. At final ($T = 2$), if the efficacy boundaries are not met, the cohort automatically stops for futility. The initial letters E or F in the superscript of the thresholds $\delta$ and $\gamma$ indicate if this boundary is used to stop for efficacy (E) or futility (F). The subscripts refer to the treatment arms (C (combination), A (monotherapy A), B (monotherapy B) and S (SoC)) and allow for different thresholds for the individual comparisons. Choosing e.g. $\gamma_{CA}^{E,1} = \gamma_{CB}^{E,1} = \gamma_{AS}^{E,1} = \gamma_{BS}^{E,1} = 1$ corresponds to not allowing early stopping for efficacy at interim. If at interim both stopping for early efficacy and futility is allowed, parameters need to be chosen carefully such that GO and STOP and decisions are not simultaneously possible. 

\subsection{Data Sharing and Allocation Ratios} \label{Allocation Ratios}

The advantage of platform trials is that they can use the data from all cohorts, which potentially increases the power in every individual cohort. We can specify whether we want to share information on the backbone monotherapy and SoC arms across the study cohorts. Several different methods have been proposed to facilitate adequate borrowing of non-concurrent (these can be internal or external to the trial) controls \citep{Viele2014, Schmidli2014, Li2020, Harun2020, Feisst2020, Bofill2021}. We consider four options, all applying to both SoC and backbone monotherapy: 1) no sharing, using only data from the current cohort (see first row of Figure \ref{fig:sharing}), 2) full sharing of all available data, i.e. using all data 1-to-1 (see second row of Figure \ref{fig:sharing}), 3) only sharing of concurrent data, i.e. using concurrent data 1-to-1 (see third row of Figure \ref{fig:sharing}) and 4) using a dynamic borrowing approach further described in appendix \ref{dynamic_borrowing}, in which the degree of shared data increases with the homogeneity of the observed data in the current cohort and the pooled observed data of all other cohorts, i.e. discounting the pooled observed data of other cohorts less, if the observed response rate is similar (see fourth row of Figure \ref{fig:sharing}). Whenever a new cohort enters the platform trial, a key design element is the allocation ratio to the newly added arms (combination therapy, add-on monotherapy, backbone monotherapy and SoC) and whether the allocation ratio of the already ongoing cohorts should be changed as well, e.g. randomizing less patients to backbone monotherapy and SoC in case this data is shared across cohorts. The platform trial advances dynamically and as a result the structure can follow many different trajectories (it is unknown in the beginning of the trial how many cohorts will enter the platform, how many of them will run concurrently, whether the generated data will stem from the same underlying distributions and should therefore be shared, etc.). As the best possible compromise under uncertainty, we aimed to achieve a balanced randomization for every comparison in case of either no data sharing or sharing only concurrent data. Depending on the type of data sharing and the number of active arms, this means either a balanced randomization ratio within each cohort in case of no data sharing (i.e. 1:1:1:1, combination : add-on mono (monotherapy B) : backbone mono (monotherapy A) : SoC), or a randomization ratio that allocates more patients to the combination and add-on monotherapy arm for every additional active cohort in case of using only concurrent data. As an example, if at any point in time $k$ cohorts are active at the same time and we share only concurrent data, the randomization ratio is $k:k:1:1$ in all cohorts, which ensures an equal number of patients per arm for every comparison. This allocation ratio is updated for all active cohorts every time the number of active cohorts $k$ changes due to dropping or adding a new cohort. As an example, assume 30:30:30:30 patients have been enrolled in cohort 1 before a second cohort is added. Then, until e.g. an interim analysis is performed in cohort 1, both cohorts will have an allocation ratio of 2:2:1:1. If the interim analysis is scheduled after 180 patients per cohort, this would mean another 20:20:10:10 patients need to be enrolled in cohorts 1 and 2, since we can use the 10 concurrently enrolled backbone monotherapy and 10 concurrently enrolled SoC patients in cohort 2 for the interim analysis in cohort 1, leading to a balanced 50:50:50:50 patients for the comparisons.  In case of either full sharing or dynamic borrowing, we use the same approach as when using concurrent data only.

\subsection{Definition of Cohort Success and Operating Characteristics} \label{OCs}

As discussed in section \ref{Trial Design}, for testing the efficacy of the combination therapy, four pair-wise comparisons are conducted. If we were running a single, independent trial investigating a combination therapy, we would consider the trial a success if all of the necessary pair-wise comparison were successful. Consequently, we would consider it a failure if at least one of the necessary pair-wise comparison was unsuccessful. For clearer understanding, we call these two options respectively a positive or negative outcome. Depending on the formulated hypotheses, these might be either true or false positives or negatives. To allow evaluation of the decision rules in terms of frequentist type 1 error and power, for each of the four pair-wise comparisons we formulate a set of hypotheses of the following sort (exemplary for combination versus monotherapy A, but analogously for all other comparisons): $H_0: \pi_C \leq \pi_A + \zeta_{CA}$ versus $H_1: \pi_C > \pi_A + \zeta_{CA}$. Please note that as we are drawing the response rates randomly for every new cohort entering the platform trial, it is unknown a priori whether this null hypothesis holds. Furthermore, please note while for our simulations this is not the case, $\zeta_{CA}$ can be different from $\delta_{CA}$ chosen in section \ref{Decision Rules} (in our simulations, for all pair-wise comparisons $\zeta = \delta = 0$). We will elaborate more on what happens in this case in appendix \ref{zeta}. If all such pair-wise alternative hypotheses within a cohort hold, we call the cohort truly efficacious (and any decision made is either a true positive or a false negative), otherwise we call it truly not efficacious (and any decision made is either a true negative or a false positive). The latter implies that this includes scenarios where for some of the pair-wise hypotheses of interest the alternative and for some the null holds. Depending on the decision rules used, it could be debated if other definitions are more meaningful (e.g. if the comparisons of the combination therapy to the monotherapies are both true alternatives the cohort may be considered as efficacious). While for a single, independent trial this would yield a single outcome (true positive, false positive, true negative, false negative), for a platform trial with multiple cohorts this yields a vector of such outcomes, one for each investigated cohort. In order to evaluate different trial designs, operating characteristics need to be chosen that take into account the special features of the trial design, but are at the same time interpretable in the classical context of hypothesis testing. In more simple trial designs evaluating one treatment against a control, power and type 1 error rates are used to judge the design under consideration. For platform trials, similarly to multi-arm multi-stage trials \citep{Bretz2010}, the choice of operating characteristics is not obvious \citep{Wason2014_2, Howard2018, Stallard2019, Collignon2020, berry2020potential}. Furthermore, for the particular trial design under consideration, many operating characteristics based on the pair-wise comparisons of the different monotherapies, SoC and combination therapy could be considered. We decided to investigate cohort-level and platform-level operating characteristics (see figure \ref{fig:trialdesign}). As described previously, when simulating platform trials we might allow the same treatment arms to have different response rates in different cohorts. Even if we allow for a mix of true null and alternative hypotheses, for a single simulated platform trial, it could happen by chance that only efficacious or only not efficacious cohorts are added. Consequently, for the platform-level operating characteristics, a definition challenge arises when by chance either no true positive and false negative (in case all cohorts are truly not efficacious) or true negative and false positive (in case all cohorts are truly efficacious) decisions are possible. To make sure that the operating characteristics reflect this situation (which depends on the prior on the treatment effects), we differentiate between counting all simulation iterations (which implicitly takes into account the prior on the treatment effect, ``BA" (``Bayesian Average") operating characteristics, FWER BA and Disj Power BA) or only those simulation iterations where a false decision could have been made towards the type 1 error rate and power (FWER and Disj Power). An overview of the operating characteristics used in this article and their definitions can be found in table \ref{tab:ocs}.


\newcolumntype{L}[1]{>{\raggedright\let\newline\\\arraybackslash\hspace{0pt}}m{#1}}
\newcolumntype{C}[1]{>{\centering\let\newline\\\arraybackslash\hspace{0pt}}m{#1}}
\newcolumntype{R}[1]{>{\raggedleft\let\newline\\\arraybackslash\hspace{0pt}}m{#1}}

\begin{longtable}{@{\extracolsep{5pt}} L{3cm}L{11cm}L{1cm}} 
  \caption{Operating characteristics used in this paper and their definitions.} 
  \label{tab:ocs} 
\\[-1.8ex]\hline 
\hline \\[-1.8ex] 
Name & Definition \\ 
\hline \\[-1.8ex] 
\hline \\[-1.8ex] 
PCP & ``Per-Cohort-Power", the ratio of the sum of true positives among the sum of truly efficacious cohorts (i.e. the sum of true positives and false negatives) across all platform trial simulations, i.e. the probability of a false positive decision for any new cohort entering the trial. This is a measure of how wasteful the trial is with superior therapies. \\
\hline \\[-1.8ex] 
PCT1ER & ``Per-Cohort-Type-1-Error", the ratio of the sum of false positives among the sum of all truly not efficacious cohorts across all platform trial simulations, i.e. the probability of a true positive decision for any new cohort entering the trial. This is a measure of how sensitive the trial is in detecting futile therapies. \\
\hline \\[-1.8ex] 
FWER & "Family-wise type 1 Error Rate", the proportion of platform trials, in which at least one false positive decision has been made (i.e. probability of at least one false positive decision across all cohorts), where only such trials are considered, which contain at least one cohort that is in truth futile. Formal definition: $\frac{1}{|I_0^{\ast}|} \sum_{i \in I_0^{\ast}} \mathbbm{1} \{ FP_i > 0 \}$, where $I_0^{\ast} = \{i \in \{1, ... iter \}: n_{i}^{H0} > 0 \}$, $iter$ is the number of platform trial simulation iterations, $FP_i$ denotes the number of false-positive decisions in simulated platform trial $i$ and $n_{i}^{H0}$ is the number of not efficacious cohorts in platform trial $i$. \\
\hline \\[-1.8ex] 
FWER BA & "Family-wise type 1 Error Rate Bayesian Average", the proportion of platform trials, in which at least one false positive decision has been made (i.e. probability of at least one false positive decision across all cohorts), regardless of whether or not any cohorts which are in truth futile exist in these trials. Formal definition: $\frac{1}{iter} \sum_{i=1}^{iter} \mathbbm{1} \{ FP_i > 0 \}$, where $iter$ is the number of platform trial simulation iterations and $FP_i$ denotes the number of false-positive decisions in simulated platform trial $i$. This will differ from FWER in scenarios where - due to a prior on the treatment effect - in some simulation runs, there are by chance only efficacious cohorts in the platform trial (see section \ref{Efficacy} for more details on the different treatment efficacy scenarios). \\
\hline \\[-1.8ex] 
Disj Power & "Disjunctive Power", the proportion of platform trials, in which at least one correct positive decision has been made (i.e. probability of at least one true positive decision across all cohorts), where only such trials are considered, which contain at least one cohort that is in truth superior. Formal definition: $\frac{1}{|I_1^{\ast}|} \sum_{i \in I_1^{\ast}} \mathbbm{1} \{ TP_i > 0 \}$, where $I_1^{\ast} = \{i \in \{1, ... iter \}: n_{i}^{H1} > 0 \}$, $iter$ is the number of platform trial simulation iterations, $TP_i$ denotes the number of true-positive decisions in simulated platform trial $i$ and $n_{i}^{H1}$ is the number of efficacious cohorts in platform trial $i$. \\
\hline \\[-1.8ex] 
Disj Power BA & "Disjunctive Power Bayesian Average", the proportion of platform trials, in which at least one correct positive decision has been made (i.e. probability of at least one true positive decision across all cohorts), regardless of whether or not any cohorts which are in truth superior exist in these trials. Formal definition: $\frac{1}{iter} \sum_{i=1}^{iter} \mathbbm{1} \{ TP_i > 0 \}$, where $iter$ is the number of platform trial simulation iterations and $TP_i$ denotes the number of true-positive decisions in simulated platform trial $i$. This will differ from FWER in scenarios where - due to a prior on the treatment effect - in some simulation runs, there are by chance no efficacious cohorts in the platform trial (see section \ref{Efficacy} for more details on the different treatment efficacy scenarios).  \\
\hline \\[-1.8ex] 
\hline \\[-1.8ex] 
\end{longtable} 


\section{Simulations} \label{Simulations}

\subsection{Simulation Setup}

We investigate the impact of a range of different design parameters and assumptions on the operating characteristics. In total, we investigated fourteen different settings with respect to the treatment efficacy assumptions for the combination arm, the monotherapy arms and the SoC arm. In the main text, with one exception, only results of one treatment efficacy setting (setting 1) are shown as in this scenario both truly efficacious and not efficacious cohorts can enter the platform trial (see section \ref{OCs}). In the investigated setting (setting 1), for every new cohort entering the platform trial, the backbone monotherapy (response rate 20\%) is superior to SoC (response rate 10\%). The add-on monotherapy efficacy is random with 50\% probability to be as efficacious as the backbone monotherapy (response rate 20\%) and 50\% probability to be not efficacious (response rate 10\%). Adding to the monotherapies, the combination therapy interaction effect is additive, meaning the combination therapy is superior to both monotherapies if the add-on monotherapy is efficacious (response rate 40\%) and not superior to the backbone monotherapy otherwise (response rate 20 \%). In terms of sample sizes, we vary the final cohort sample size from 100-500 in steps of 100 and fix the interim sample size at half of the final sample size. The maximum number of cohorts entering the platform trial over time is varied between 3 and 7 in steps of 2 and the probability to include new cohorts in the platform after every patient is set to either 1\% or 3\%. For the Bayesian decision rules (see section \ref{Decision Rules}), by default we set all $\delta = 0$, all $\gamma^{E,T} = 0.9$ and all $\gamma^{F,T} = 0.5$ in equation \ref{eq:DR}. An overview of the simulation setup can be found in table \ref{tab:simsetup}. It should be obvious that the chosen decision rules will yield dramatically different power and type 1 errors across different simulation parameters and treatment efficacy settings. It was our primary goal to investigate the relative impact of the simulation parameters and treatment efficacy settings on the operating characteristics. Of course, for any given combination of simulation parameters and treatment efficacy assumptions, the decision rules could be adapted to e.g. achieve a per-cohort power of 80\%. We will investigate the impact of the decision rules in more detail in section \ref{Impact of Decision Rules}. \newline
\newline
For every configuration of design parameters and assumptions, 10000 platform trials were simulated. Unless otherwise specified, simulation results presented in this article use treatment effect scenario 1, the above mentioned Bayesian decision rules, a final sample size of 500 per cohort, a maximum number of cohorts per platform trial of up to 7 and a probability of including a new cohort after every patient of 3\%. Results of all further treatment efficacy assumptions, as well as a table summarizing all different treatment efficacy settings are presented in the supplements (table \ref{tab:settings}). For a detailed overview of the general simulation assumptions, as well as all possible simulation parameters for the R software package and Shiny App, see the R package \textbf{CohortPlat} vignette on \href{https://el-meyer.github.io/CohortPlat/articles/my-vignette.html}{GitHub}. The R software package can be downloaded from \href{https://el-meyer.github.io/CohortPlat/}{GitHub} or CRAN. Further results not discussed in the main text can be found in the supplements. For a complete overview of all simulation results, we developed a Shiny App facilitating self-exploration of all of our simulation results. The purpose of R Shiny app is to quickly inspect and visualize all simulation results that were computed for this paper. We uploaded the R Shiny app, alongside all of our simulation results used in this paper to \href{https://sny.cemsiis.meduniwien.ac.at/~zrx5rdf/oWer32/}{our server} \citep{Shinyserver}.

\begin{longtable}{@{\extracolsep{5pt}} L{2cm}L{2cm}L{2cm}L{8cm}} 
\caption{Simulation Setup Overview. For different simulation parameters, we differentiate between parameters that are considered a design choice and parameters that are considered an assumption regarding the future course of the platform or treatment effects. For the investigated values, we state in bold which value was considered the default value, i.e. unless stated otherwise in a particular figure, the parameters was set to this value. For some parameters there is no default (e.g. when shown as a simulation dimension in every figure or if chosen as a fixed design parameter in section \ref{Methods}).}
  \label{tab:simsetup}
  \\[-1.8ex]\hline 
\hline \\[-1.8ex] 
Name & Type & Investigated Values & Description \\
\\[-1.8ex]\hline 
\hline \\[-1.8ex] 

Maximum Number of Cohorts  & Assumption &  3, 5, \textbf{7}  & Assumed maximum number of cohorts per platform (can be less in individual simulations) \\
\hline

Cohort Inclusion Rate  & Assumption &  0.01,  \textbf{0.03}  & Probability to include a new cohort in the ongoing platform trial after every simulated patient (unless maximum number of cohorts is reached). The default value leads to reaching the assumed maximum number of cohorts in nearly every simulation for nearly every sample size. \\
\hline

Treatment Efficacy Setting  & Assumption & \textbf{1}, 2-14 & Assumed treatment effects of the different study arms. For more information, see table \ref{tab:settings}. \\
\hline

Final Cohort Sample Size & Design Choice & 100, 200, 300, 400, \textbf{500} & Number of patients after which final analysis in a cohort is conducted (interim analysis always after half the final sample size) \\
\hline

Data Sharing & Design Choice & all, concurrent, dynamic, cohort & Different methods of data sharing used at analyses, ranging from full pooling to not sharing at all. For more information, see section \ref{Allocation Ratios}. \\
\hline

Allocation Ratios & Design Choice & Balanced or Unbalanced (depends on data sharing) & Allocation ratio to treatment arms within a cohort. Depending on the method of data sharing, the allocation ratio is set to either balanced or unbalanced (i.e. randomizing more patients to combination and add-on monotherapy). For more information, see section \ref{Allocation Ratios}. \\
\hline

Bayesian Decision Rule & Design Choice & $\delta \in [\textbf{0}, 0.2]$,  $\gamma \in [0.65, 0.95]$ \ (\textbf{0.9}) & Thresholds used in the Bayesian decision making at interim and final. Values other than the default values are used only in figure \ref{fig:impact_dec}. For more information, see section \ref{Decision Rules}. \\
\hline

Interim Analysis & Design Choice & \textbf{Early Stopping for Futility}, No Early Stopping for Futility & Binding rules on whether or not a cohort can be stopped at interim for early futility (note: cohorts can always stop for early efficacy). For more information, see section \ref{Decision Rules}. Further extensions investigated in the supplements include using a surrogate short-term endpoint for interim decision making. \\
\hline

\end{longtable}

\subsection{Simulation Results} \label{Results}

\begin{sidewaysfigure}[ht]
\includegraphics[scale=0.45]{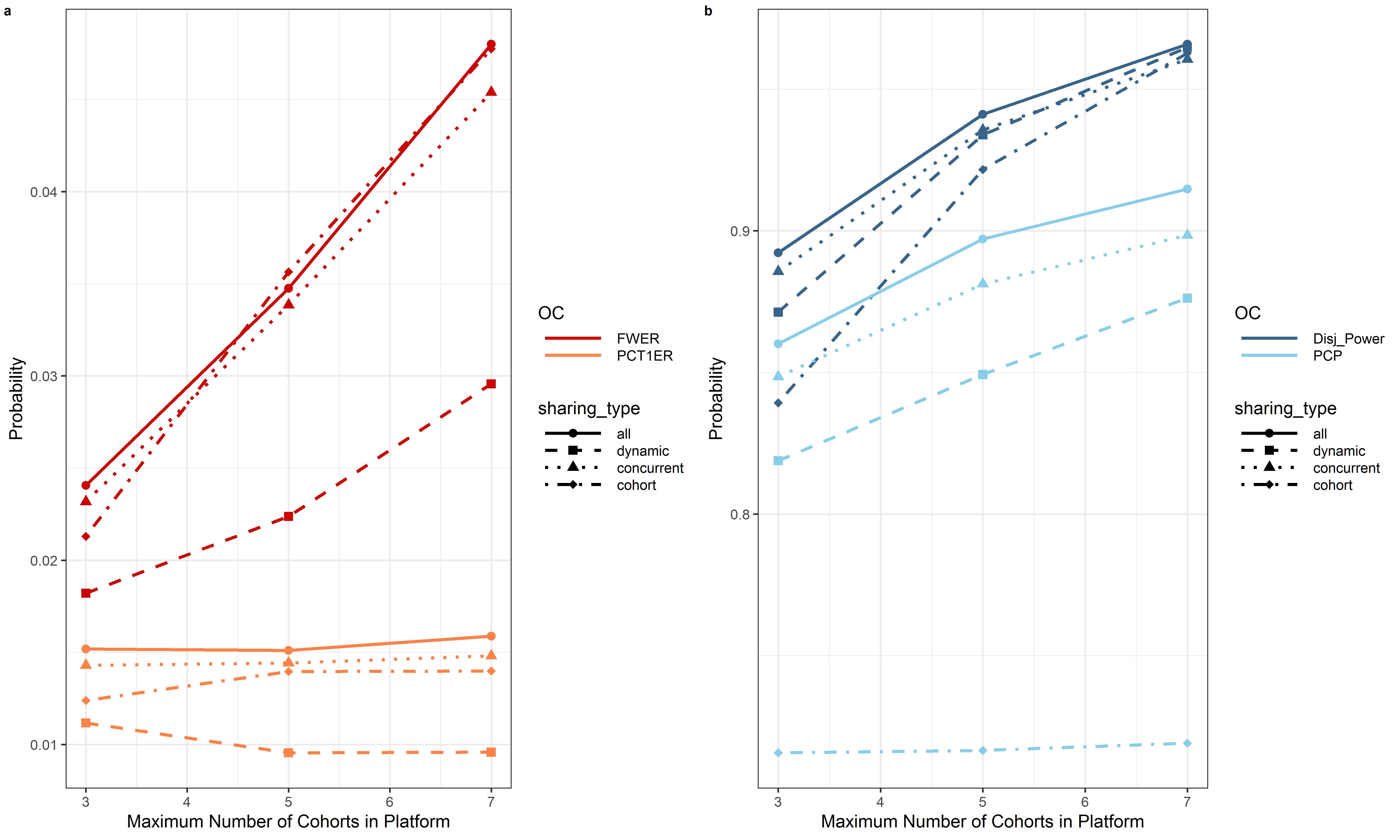}
\caption{Impact of the data sharing (linetype and point shape) and maximum number of cohorts per platform (x-axis) on the per-cohort and per-platform type 1 error (figure \ref{fig:g33}a) and power (figure \ref{fig:g33}b) in treatment efficacy setting 1. Please note that different scaling of the y-axis is used in the two subfigures. With increasing number of cohorts in the platform, the chance to make at least one correct positive or negative decision increases. When data is shared, the per-cohort power increases, while it stays constant when no data sharing is planned.}
\label{fig:g33}
\end{sidewaysfigure}

\begin{sidewaysfigure}[ht]
\includegraphics[scale=0.45]{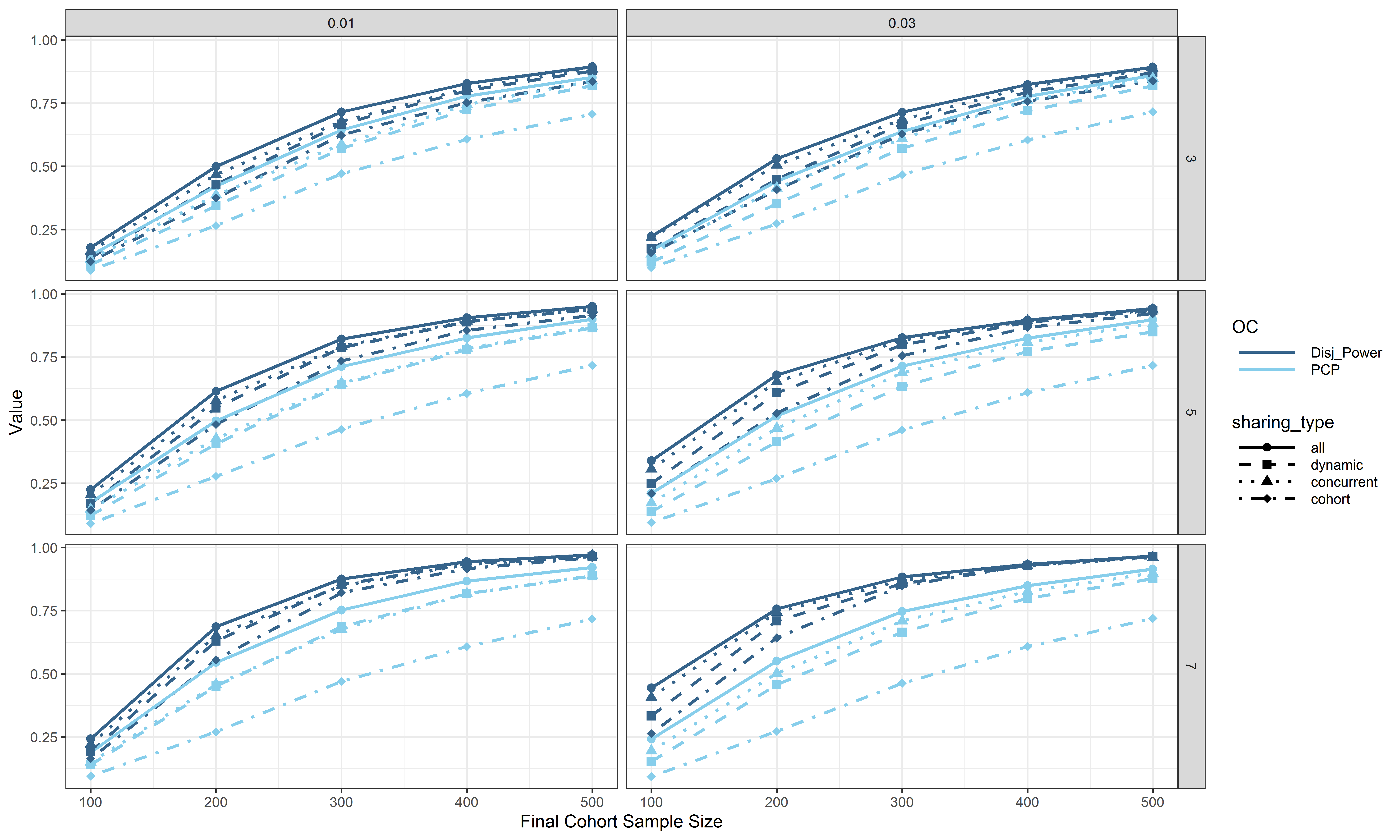}
\caption{Per-cohort and per-platform power with respect to data sharing (linetype and point shape), assumptions regarding the maximum number of cohorts (rows), cohort inclusion rate (columns) and final cohort sample size (x-axis) in treatment efficacy setting 1. Generally, both types of power increase with increasing final cohort sample size. In case of no data sharing, the per-cohort power is independent of assumptions regarding the expected platform size and cohort inclusion rate.}
\label{fig:g31_SUP_power_complexity}
\end{sidewaysfigure}

In figure \ref{fig:g33}, we investigate the impact of the data sharing and maximum number of cohorts per platform on PCT1ER and FWER (figure \ref{fig:g33}a) and PCP and disjunctive power (figure \ref{fig:g33}b). We observe that the PCP does not increase with the platform size (maximum number of cohorts) in case of no data sharing, while it does increase with the magnitude of data sharing. For the disjunctive power we observe a similar relationship, i.e. it increases regardless of the data sharing with increasing number of investigated cohorts. Furthermore, we see that the PCT1ER stays more or less constant with respect to the platform size, however is lowest for the dynamic borrowing approach. We believe that this is due to the following: A type 1 error in a cohort happens only if either the observed combination therapy and add-on monotherapy efficacy are on a random high or if the observed backbone and SoC response rates are on a random low. In the former case, the degree of data sharing will have no impact on the decisions made. In the latter case, if dynamic borrowing is used, all other cohorts will discount this data, leading to fewer type 1 errors in the other cohorts compared to if complete pooling was used. On the other hand, for the cohort that had observed backbone and SoC response rates on a random low, there is a chance to not make a type 1 error when any sort of data sharing is used. As a result of the PCT1ER behaviour, the FWER increases with increasing platform size and is again lowest in the case of dynamic borrowing.
\newline
\newline
In figure \ref{fig:g31_SUP_power_complexity}, we investigate the impact of optimistic/pessimistic assumptions regarding the expected platform size, cohort inclusion rate and final cohort sample size on per-cohort and per-platform power. We observe that the PCP is independent of the assumptions regarding the maximum number of included cohorts and the cohort inclusion rate. When the sample size is increased, both PCP and disjunctive power increase, however the increase is faster when using more optimistic assumptions regarding the expected platform size and cohort inclusion rate, as firstly more data can be shared and secondly more cohorts will be included on average, thereby increasing the chance for any true positive decision. \newline
\newline
Next, we removed the 50:50 chance for new cohorts entering the platform trial to have an efficacious add-on monotherapy and increased this probability to 100\% (i.e. we are using treatment efficacy setting 7 instead of setting 1). We also wanted to investigate the impact of an increase in SoC response rate, so we increased the SoC response rate from 10\% to 20\% but kept the incremental increases in response rate the same (i.e. 10\% points increased for the monotherapies and 30\% points increased for the combination therapy; setting 14). Results for the power are presented in the supplements in figure \ref{fig:g61}. While all the relative influences of level of data sharing, use of decision rules, etc. appear to be unchanged, we observed consistently slightly lower power for increased SoC response rate, which could be due to increased variance in the observed response rates. Another phenomenon that is particularly pertinent in this figure: While the PCP is always lowest when sharing no data, the disjunctive power is only the lowest when sharing no data when using small sample sizes. With increasing sample size the disjunctive power is greatest when sharing no data. While this might seem not intuitive at first, we believe it is due to random highs in the SoC (leading to unsuccessful comparisons of monotherapies versus SoC) and backbone monotherapy (leading to unsuccessful comparisons of the combination therapy versus backbone monotherapy) arms. When these occur, sharing this data across cohorts might negatively impact all other truly efficacious cohorts leading to simultaneously only false negative decisions. As a result, the disjunctive power drops. This cannot happen when no data is shared. 


\begin{sidewaysfigure}[ht]
\includegraphics[scale=0.45]{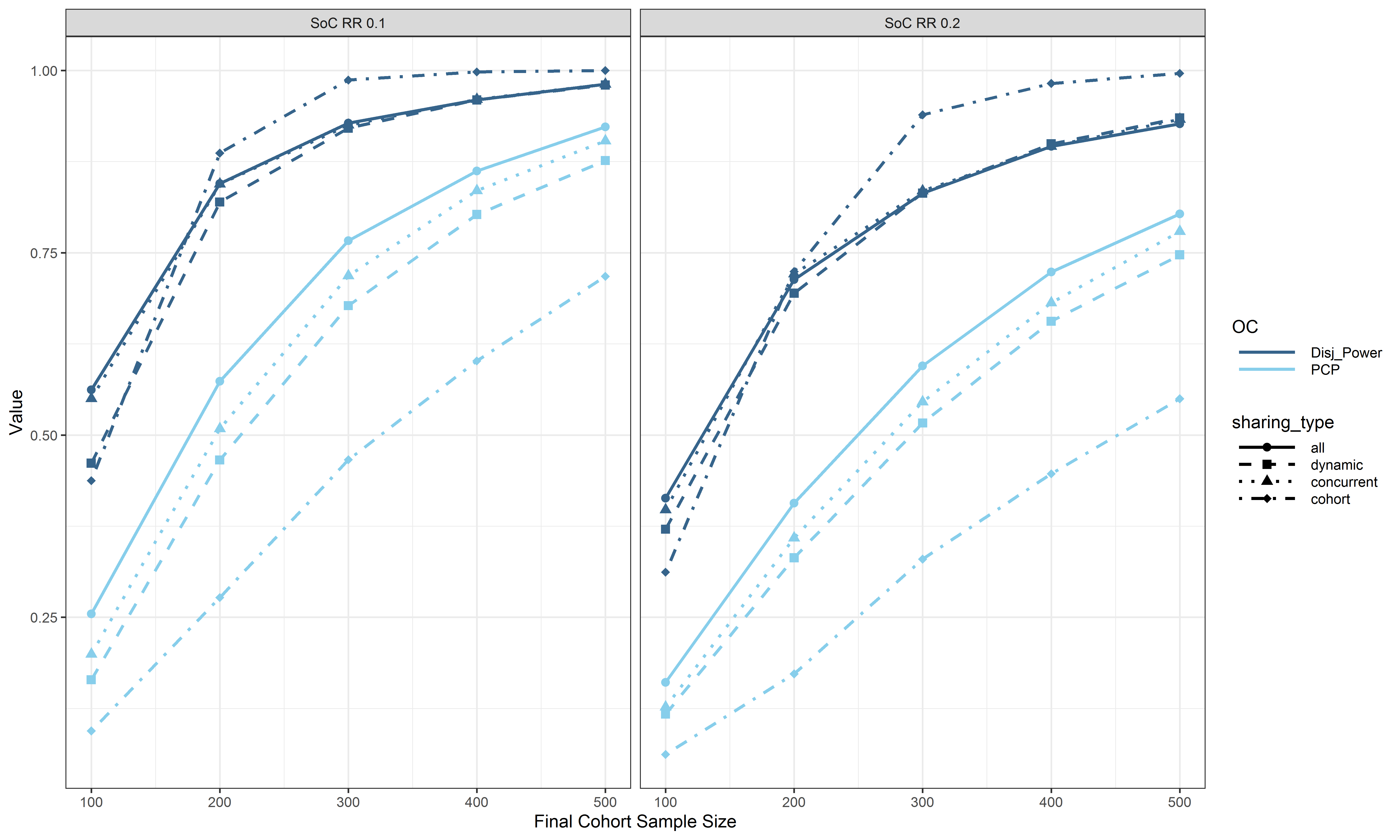}
\caption{Impact of SoC response rate (columns; treatment efficacy settings 7 versus 14), final cohort sample size (x-axis) and data sharing (linetypes and point shapes) on per-cohort and per-platform power. We observed consistently lower power for increased SoC response rate. We also observed that while for lower sample sizes the per-platform power is increased with increasing amount of data sharing, this is not true anymore for larger sample sizes.}
\label{fig:g61}
\end{sidewaysfigure}


\subsubsection{Impact of Modified Decision Rules} \label{Impact of Decision Rules}

\begin{sidewaysfigure}[ht]
\includegraphics[scale=0.4]{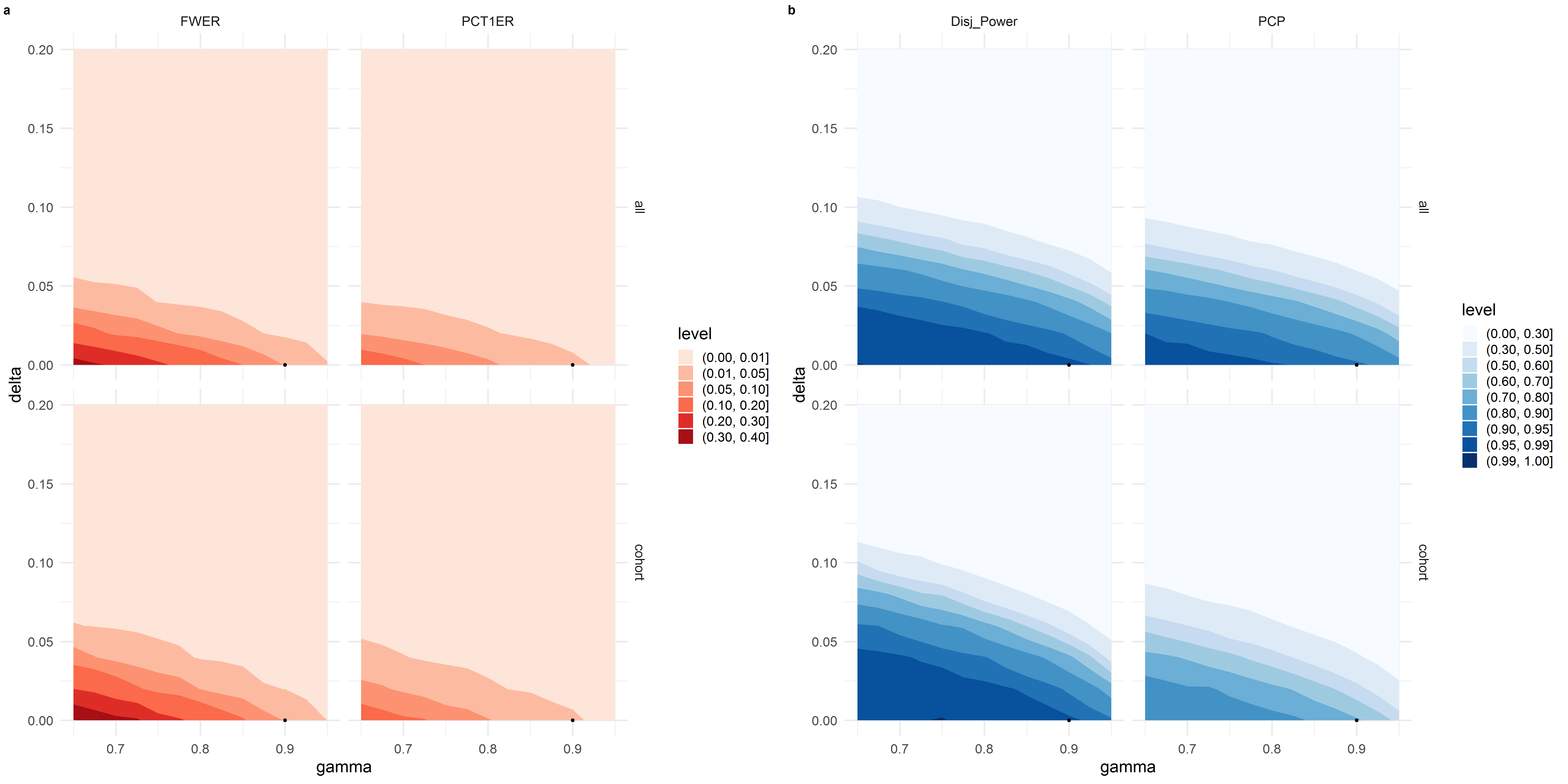}
\caption{Impact of the Bayesian GO decision rules on a) per-cohort and per-platform type 1 error rates and b) per-cohort and per-platform power in treatment efficacy setting 1. The black dot corresponds to the decision rules chosen in section \ref{Decision Rules}. For every error rate, we set the data sharing (rows) to either full ("all") or none ("cohort"). The x-axis shows the required confidence (``gamma'') and the y-axis the required superiority (``delta'') used in the Bayesian decision making in section \ref{Decision Rules}. It is apparent that by choice of $\delta$ and $\gamma$, a wide range of different error rates can be obtained.}

\label{fig:impact_dec}
\end{sidewaysfigure}

We further investigated the impact of parameters required for the Bayesian GO decision rules on the error rates. Every individual comparison includes a superiority margin $\delta$ and a required confidence $\gamma$ (e.g. posterior $(P(\pi_{Comb} > \pi_{MonoA} + \delta | Data) > \gamma))$. In the decision rules in section \ref{Decision Rules} we fixed $\delta = 0$ and $\gamma = 0.90$. We now varied $\gamma$ from 0.65 to 0.95 and $\delta$ from 0 to 0.2. The impact on the operating characteristics is presented in figure \ref{fig:impact_dec}. Figure \ref{fig:impact_dec}a reveals that both when sharing no or all data, the FWER can increase beyond 0.05 and even far beyond 0.10. Similarly, figure \ref{fig:impact_dec}b reveals that when choosing more lenient decision rules and increasing the data sharing, a PCP of close to 1 is attainable. Such contour plots will help to fine tune the Bayesian decision rules in order to achieve the desired operating characteristics. For $\gamma$, one would usually expect values equal to or larger than 80\%. The choice of $\delta$ also relies on clinical judgment. Generally, as a result of the combined decision rule, operating characteristics are rather conservative with respect to type 1 error rates. 


\subsubsection{Summary of Main Results}

Planning a platform trial is much more complex than planning a traditional full development trial. Firstly, both power and type 1 error might have to be controlled on the platform or cohort level. A platform trial that replaces sponsors' individual trials with the aim of advancing as many compounds as possible will try to control error rates on the cohort level. However, a platform trial with the goal of making sure that at least one efficacious compound is advanced would rather control error rates on the platform level. As seen in figure \ref{fig:g31_SUP_power_complexity}, the per-cohort power depends on the assumptions regarding the cohort inclusion rate and number of cohorts that will enter over time. In case there is substantial uncertainty regarding how fast and how many new cohorts will enter the platform trial over time when determining the sample size, conservative lower bounds for the per-cohort power can be used by assuming no data sharing will take place. In this case, further simulations revealed that a final cohort sample size of around 600 would be needed to achieve a PCP of 0.8. In order to achieve a disjunctive power of 0.8, a final cohort sample size in the range of 300-500 is needed (depending on the assumptions; for few cohorts and a lower probability of cohort inclusion roughly 500 and for more cohorts and a higher probability if inclusion slightly below 300). In the fortunate event that the platform will attract many cohorts, the power for future cohorts would only increase with data sharing while guaranteeing a certain power for the first cohorts. If the priority is PCP and we are extremely confident about new cohorts entering the platform over time, by simply pooling all data we might achieve a PCP of 0.8 with a final cohort sample size of around 340 and a disjunctive power of 0.8 with a final cohort sample size of around 220. However, the above results are only true for the chosen treatment efficacy assumptions. When we are more optimistic about the treatment efficacy, the best choice for maximizing disjunctive power might be to not share any data at all (as seen in figure \ref{fig:g61}, where in the left panel a disjunctive power of 0.8 is achieved with a final cohort sample size below 200 and no data sharing). 

\section{Discussion} \label{Discussion}

To our knowledge, we conducted the first cohort platform simulation study to evaluate combination therapies with an extensive range of simulation parameters that reflect the potential complexity and a priori unknown trajectory of platform trials. The design under investigation is an open-entry, cohort platform study with a binary endpoint evaluating the efficacy of a number of two-compound combination therapies with a common backbone monotherapy compared to the respective monotherapies with putative individual efficacy over SoC. For example, the SoC arm could include a placebo add-on to achieve blinding. A range of treatment effect scenarios, as well as types of data sharing were investigated, from no sharing to full sharing. In our simulations, no a priori knowledge is required regarding the possible trajectories of the platform trials, as in every simulation, different trajectories are dynamically simulated based on randomly generated events such as patients, outcomes, cohort inclusions, etc. This means in one simulation, the platform trial might be stopped after including only one cohort and in another simulation run, it might stop after evaluating 7 cohorts. This is also a major difference compared to simulation programs for multi-arm multi-stage studies, which usually assume a fixed number of different arms or compute only arm-wise operating characteristics. Within each simulation in our program, periods of overlap between cohorts are recorded. We believe that especially this dynamic simulation is one of the key strengths of our simulation study, as it produces different realistic trajectories and computes operating characteristics across all such possible trajectories. We developed an R package and Shiny app alongside that facilitate easy usage and reproducibility of all simulation code, while the Shiny app for result exploration facilitates reproduction of our extensive simulation results, which we could not all discuss in this paper. Development of software for platform trial simulation is not straightforward and time spent on programming increases rapidly when more realistic features such as staggered entry, adding and dropping of treatment arms or data sharing across treatments/cohorts are included \citep{Meyer2021}. \newline
\newline
Our results indicate that - apart from the treatment efficacy assumptions - the method of data sharing, exact specification of decision rules and complexity of the platform trial (expressed as maximum number of cohorts and cohort inclusion rate) are most influential to the operating characteristics of the platform trial. As expected, in nearly all cases, pooling all data leads to the largest power and type 1 error rates, whereas no data sharing leads to the lowest power. Methods that do not pool all data, but either discount them or use only concurrently enrolled patients, might significantly increase the power while only marginally increasing or even decreasing the type 1 error. Furthermore, definition of error rates in the context of a cohort platform trial with a combined decision rule per cohort is not straightforward \citep{Stallard2019, Collignon2020, bai2020multiplicity, parker2020non, bretz2020commentary, posch2020commentare}. As an example, whether or not to include simulated platforms that - as a result of random sampling of treatment efficacy from a specified prior distribution - do not contain truly efficacious cohorts in the calculation of the disjunctive power can lead to discrepancies of up to 15\% points in our simulations. In terms of type 1 error rates, we defined a type 1 error via a target product profile and focused on interpreting the per-cohort type 1 error and the family-wise error rate in this paper. Other authors have recently suggested that FDR control might yield better properties than family-wise error rate control in the multi-arm context \citep{Wason2020}. The control of error rates such as the FDR might be appropriate for exploratory platform trials, which are less of a regulatory concern. Furthermore, in our investigations we treat configurations of "partial" superiority (e.g. the case where the combination is declared superior to the two monotherapies, but one or both of these are not declared superior to SoC) as non-rejections of the null hypothesis. Reversely, when for any of the four pair-wise comparisons conducted the null hypothesis holds, we consider any successful decision on the cohort level a false positive decision. For some configurations of null and alternative hypotheses within a cohort (e.g. when the combination therapy is superior to both monotherapies and SoC, but one of the components is not superior to SoC), the per-cohort type 1 error and the family-wise error rate are much higher than in the global null settings. For these configurations, partial power concepts might be more appropriate, or more relaxed testing strategies or thresholds than the one implemented could be considered, e.g. if superiority of the backbone over SoC does not have to be shown within the platform trial. In this case, three instead of four pairwise comparisons would be required. Also it might make sense to use more relaxed thresholds for the comparisons of monotherapies versus SoC than for the comparisons of combination therapy versus the monotherapies. A further simplification would be to only test the combination therapy against the monotherapies (two pairwise comparisons) or against SoC (one pairwise comparison), in which case also the number of arms per cohort and allocation ratio would have to be adapted. For example in oncology, sometimes it is sufficient to demonstrate synergistic effects of the combination therapy, because monotherapies cannot separate from SoC. Such a combination therapy trial would be a success if combination therapy is superior to both monotherapies and SoC, even if the monotherapies do not separate from SoC.
\newline
\newline
As for any simulation program, especially in a highly complex and dynamic context such as platform trials, we had to make simplifications. The following restrictions must be acknowledged: 1) Patients are simulated in batches to achieve the targeted allocation rate. In more detail, for every time interval a batch of patients is simulated, all entering the study at the same time (imitating a block randomization). This led to potentially slightly different sample sizes at interim and final than planned. However, this is also not uncommon in reality. By simulating the outcomes this way, the overall run time could be significantly reduced, thereby allowing us to investigate more than ten thousand combinations of simulation parameters and simulating more than one hundred million individual platform trajectories. For more information on the exact patient sampling mechanism, see the R package vignette. 2) The dynamic borrowing approach we used is rather straightforward and heuristic. While based on a well known and accepted method described by \citet{Schmidli2014}, we had to adapt the approach to be compatible with our cohort platform design and simulation structure, leading to some heuristic adaptations for the sake of computational efficiency. 3) Furthermore, in the simulations individual patient demographic or baseline characteristics were not considered. We assumed that patients were always drawn from the same population and different cohorts shared common inclusion/exclusion criteria. All newly simulated patients under such assumptions were eligible for randomization to any cohort ongoing at the time the patient was simulated, which enabled naive borrowing of data across cohorts during the analyses. In practice, separate cohorts could have some different inclusion / exclusion criteria due to scientific (e.g. treatments of different modes of action) or operational considerations. Nevertheless, the schematic approaches in this simulation study for assessing trial operating characteristics could remain instrumental. Otherwise, partial instead of complete data sharing could be planned for when designing a platform study. In such practical scenarios, prior epidemiological and clinical knowledge on the disease of interest also plays a critical role for appropriate data borrowing hence for the reliable assessment of power and error rates. 4) As we would expect in real life, our simulations show variations in the final sample size. Especially when performing more data sharing, the originally planned final sample size tends to be exceeded. Further simplifications included not simulating variations in recruitment speed due to availability of centers, external and internal events such as approval of competitor drugs or discontinuation of drug development programs. While the software does facilitate stopping treatments due to safety events, we did not use this parameter in the presented simulation study. As the best possible compromise under uncertainty, we aimed to achieve a balanced randomization for every comparison in case of either no data sharing or sharing only concurrent data and refrained from investigating response-adaptive randomization, which might increase the efficiency of the platform trial \citep{Saville2016, Viele2020}. We only investigated platform trials with a maximum number of cohorts between 3-7. We believe this is sufficient in the initial planning step. If less cohorts are expected, a platform trial is not warranted. If more cohorts are expected to be included over time, simulations would need to be re-run. \newline
\newline
To evaluate combination therapies the proposed platform design will be an efficient way to evaluate potential drugs and the resulting therapies. When exploring Bayesian decision rules, a key factor is fine-tuning of decision parameters (e.g. $\delta$ and $\gamma$) in case operating characteristics should be controlled at a certain level. In the Bayesian decision rules, we tried to demonstrate superiority for the comparison of the monotherapies versus SoC. However, in some therapeutic areas, SoC could be an active drug with a different mechanism of action. In such instances, it could be sufficient to demonstrate non-inferiority for some of the pairwise comparisons using pre-specified non-inferiority margins $\delta < 0$. Furthermore, if drugs have more than one dose level, the platform trial can be extended to allow for more dose levels per cohort. This may result in even more complex models and decision rules, depending on which data sharing models are to be used when calculating the posteriors. Considering dose-response relationships in combination trials will be a part of future research, as well as investigation of non-inferiority decision rules and impact of different allocation ratios on the operating characteristics.


\subsection*{Funding}

EU-PEARL (EU Patient-cEntric clinicAl tRial pLatforms) project has received funding from the Innovative Medicines Initiative (IMI) 2 Joint Undertaking (JU) under grant agreement No 853966. This Joint Undertaking receives support from the European Union’s Horizon 2020 research and innovation programme and EFPIA and Children’s Tumor Foundation, Global Alliance for TB Drug Development non-profit organisation, Springworks Therapeutics Inc. This publication reflects the authors’ views. Neither IMI nor the European Union, EFPIA, or any Associated Partners are responsible for any use that may be made of the information contained herein. The PhD research of Elias Laurin Meyer was funded until 11/2020 by Novartis through the University and not at an individual level.

\subsection*{Competing Interests}

The research of ELM was funded by Novartis on the university and not an individual level. PM, CD-B, EG, YL are employees of Novartis Pharmaceuticals Corporation. FK reports grants on a university level from Novartis Pharma AG during the conduct of the study and from Merck KGaA outside the submitted work.

\subsection*{Acknowledgements}

The authors thank Constantin Kumaus for his support in programming the R Shiny App for online result exploration.


\footnotesize
\bibliography{references}

\begin{thebibliography}{48}
\providecommand{\natexlab}[1]{#1}
\providecommand{\url}[1]{\texttt{#1}}
\providecommand{\urlprefix}{URL }
\expandafter\ifx\csname urlstyle\endcsname\relax
  \providecommand{\doi}[1]{doi:\discretionary{}{}{}#1}\else
  \providecommand{\doi}{doi:\discretionary{}{}{}\begingroup
  \urlstyle{rm}\Url}\fi
\providecommand{\eprint}[2][]{\url{#2}}
\providecommand{\BIBand}{and}
\providecommand{\bibinfo}[2]{#2}
\ifx\xfnm\undefined \def\xfnm[#1]{\unskip,\space#1}\fi
\makeatletter\def\@biblabel#1{#1.}\makeatother
\bibitem[{Kunz et~al.(2020)Kunz, Jörgens, Bretz, Stallard, Lancker, Xi, Zohar,
  Gerlinger and Friede}]{Kunz2020}
\bibinfo{author}{Kunz CU}, \bibinfo{author}{Jörgens S}, \bibinfo{author}{Bretz
  F}, \bibinfo{author}{Stallard N}, \bibinfo{author}{Lancker KV},
  \bibinfo{author}{Xi D}, \bibinfo{author}{Zohar S}, \bibinfo{author}{Gerlinger
  C}, \bibinfo{author}{Friede T}.
\newblock \bibinfo{title}{Clinical trials impacted by the covid-19 pandemic:
  Adaptive designs to the rescue?}
\newblock \emph{\bibinfo{journal}{Statistics in Biopharmaceutical Research}}
  \bibinfo{year}{2020};\bibinfo{volume}{0}(\bibinfo{number}{ja}):\bibinfo{pages}{1--41}.
\newblock DOI:\bibinfo{doi}{10.1080/19466315.2020.1799857}.
\newblock \eprint{https://doi.org/10.1080/19466315.2020.1799857};
  \urlprefix\url{https://doi.org/10.1080/19466315.2020.1799857}.
\bibitem[{Stallard et~al.(2020{\natexlab{a}})Stallard, Hampson, Benda,
  Brannath, Burnett, Friede, Kimani, Koenig, Krisam, Mozgunov, Posch, Wason,
  Wassmer, Whitehead, Williamson, Zohar and Jaki}]{Stallard2020Covid}
\bibinfo{author}{Stallard N}, \bibinfo{author}{Hampson L},
  \bibinfo{author}{Benda N}, \bibinfo{author}{Brannath W},
  \bibinfo{author}{Burnett T}, \bibinfo{author}{Friede T},
  \bibinfo{author}{Kimani PK}, \bibinfo{author}{Koenig F},
  \bibinfo{author}{Krisam J}, \bibinfo{author}{Mozgunov P},
  \bibinfo{author}{Posch M}, \bibinfo{author}{Wason J},
  \bibinfo{author}{Wassmer G}, \bibinfo{author}{Whitehead J},
  \bibinfo{author}{Williamson SF}, \bibinfo{author}{Zohar S},
  \bibinfo{author}{Jaki T}.
\newblock \bibinfo{title}{Efficient adaptive designs for clinical trials of
  interventions for covid-19}.
\newblock \emph{\bibinfo{journal}{Statistics in Biopharmaceutical Research}}
  \bibinfo{year}{2020}{\natexlab{a}};\bibinfo{volume}{0}(\bibinfo{number}{0}):\bibinfo{pages}{1--15}.
\newblock DOI:\bibinfo{doi}{10.1080/19466315.2020.1790415}.
\newblock \eprint{https://doi.org/10.1080/19466315.2020.1790415};
  \urlprefix\url{https://doi.org/10.1080/19466315.2020.1790415}.
\bibitem[{Dodd et~al.(2020)Dodd, Follmann, Wang, Koenig, Korn, Schoergenhofer,
  Proschan, Hunsberger, Bonnett, Makowski et~al.}]{Dodd2020}
\bibinfo{author}{Dodd LE}, \bibinfo{author}{Follmann D}, \bibinfo{author}{Wang
  J}, \bibinfo{author}{Koenig F}, \bibinfo{author}{Korn LL},
  \bibinfo{author}{Schoergenhofer C}, \bibinfo{author}{Proschan M},
  \bibinfo{author}{Hunsberger S}, \bibinfo{author}{Bonnett T},
  \bibinfo{author}{Makowski M}, et~al.
\newblock \bibinfo{title}{Endpoints for randomized controlled clinical trials
  for covid-19 treatments}.
\newblock \emph{\bibinfo{journal}{Clinical Trials}}
  \bibinfo{year}{2020};\bibinfo{volume}{17}(\bibinfo{number}{5}):\bibinfo{pages}{472--482}.
\bibitem[{Horby et~al.(2020)Horby, Mafham, Bell, Linsell, Staplin, Emberson,
  Palfreeman, Raw, Elmahi, Prudon et~al.}]{Horby2020}
\bibinfo{author}{Horby PW}, \bibinfo{author}{Mafham M}, \bibinfo{author}{Bell
  JL}, \bibinfo{author}{Linsell L}, \bibinfo{author}{Staplin N},
  \bibinfo{author}{Emberson J}, \bibinfo{author}{Palfreeman A},
  \bibinfo{author}{Raw J}, \bibinfo{author}{Elmahi E}, \bibinfo{author}{Prudon
  B}, et~al.
\newblock \bibinfo{title}{Lopinavir--ritonavir in patients admitted to hospital
  with covid-19 (recovery): a randomised, controlled, open-label, platform
  trial}.
\newblock \emph{\bibinfo{journal}{The Lancet}}
  \bibinfo{year}{2020};\bibinfo{volume}{396}(\bibinfo{number}{10259}):\bibinfo{pages}{1345--1352}.
\bibitem[{Angus et~al.(2020)Angus, Derde, Al-Beidh, Annane, Arabi, Beane, van
  Bentum-Puijk, Berry, Bhimani, Bonten et~al.}]{Angus2021}
\bibinfo{author}{Angus DC}, \bibinfo{author}{Derde L},
  \bibinfo{author}{Al-Beidh F}, \bibinfo{author}{Annane D},
  \bibinfo{author}{Arabi Y}, \bibinfo{author}{Beane A}, \bibinfo{author}{van
  Bentum-Puijk W}, \bibinfo{author}{Berry L}, \bibinfo{author}{Bhimani Z},
  \bibinfo{author}{Bonten M}, et~al.
\newblock \bibinfo{title}{Effect of hydrocortisone on mortality and organ
  support in patients with severe covid-19: the remap-cap covid-19
  corticosteroid domain randomized clinical trial}.
\newblock \emph{\bibinfo{journal}{Jama}}
  \bibinfo{year}{2020};\bibinfo{volume}{324}(\bibinfo{number}{13}):\bibinfo{pages}{1317--1329}.
\bibitem[{Macleod and Norrie(2021)}]{Macleod2021}
\bibinfo{author}{Macleod J}, \bibinfo{author}{Norrie J}.
\newblock \bibinfo{title}{Principle: a community-based covid-19 platform
  trial}.
\newblock \emph{\bibinfo{journal}{The Lancet Respiratory Medicine}}
  \bibinfo{year}{2021};\bibinfo{volume}{9}(\bibinfo{number}{9}):\bibinfo{pages}{943--945}.
\bibitem[{Woodcock and LaVange(2017)}]{Woodcock2017}
\bibinfo{author}{Woodcock J}, \bibinfo{author}{LaVange LM}.
\newblock \bibinfo{title}{{Master Protocols to Study Multiple Therapies,
  Multiple Diseases, or Both.}}
\newblock \emph{\bibinfo{journal}{New England Journal of Medicine}}
  \bibinfo{year}{2017};\bibinfo{volume}{377}(\bibinfo{number}{1}):\bibinfo{pages}{62--70}.
\newblock DOI:\bibinfo{doi}{10.1056/NEJMra1510062}.
\bibitem[{Angus et~al.(2019)Angus, Alexander, Berry, Buxton, Lewis, Paoloni,
  Webb, Arnold, Barker, Berry et~al.}]{Angus2019}
\bibinfo{author}{Angus DC}, \bibinfo{author}{Alexander BM},
  \bibinfo{author}{Berry S}, \bibinfo{author}{Buxton M}, \bibinfo{author}{Lewis
  R}, \bibinfo{author}{Paoloni M}, \bibinfo{author}{Webb SA},
  \bibinfo{author}{Arnold S}, \bibinfo{author}{Barker A},
  \bibinfo{author}{Berry DA}, et~al.
\newblock \bibinfo{title}{Adaptive platform trials: definition, design, conduct
  and reporting considerations}.
\newblock \emph{\bibinfo{journal}{Nature Reviews Drug Discovery}}
  \bibinfo{year}{2019};\bibinfo{volume}{12}(\bibinfo{number}{18}):\bibinfo{pages}{797--807}.
\newblock DOI:\bibinfo{doi}{10.1038/s41573-019-0034-3}.
\bibitem[{Park et~al.(2019)Park, Siden, Zoratti, Dron, Harari, Singer, Lester,
  Thorlund and Mills}]{Park2019}
\bibinfo{author}{Park JJ}, \bibinfo{author}{Siden E}, \bibinfo{author}{Zoratti
  MJ}, \bibinfo{author}{Dron L}, \bibinfo{author}{Harari O},
  \bibinfo{author}{Singer J}, \bibinfo{author}{Lester RT},
  \bibinfo{author}{Thorlund K}, \bibinfo{author}{Mills EJ}.
\newblock \bibinfo{title}{Systematic review of basket trials, umbrella trials,
  and platform trials: a landscape analysis of master protocols}.
\newblock \emph{\bibinfo{journal}{Trials}}
  \bibinfo{year}{2019};\bibinfo{volume}{20}(\bibinfo{number}{1}):\bibinfo{pages}{1--10}.
\newblock DOI:\bibinfo{doi}{10.1186/s13063-019-3664-1}.
\bibitem[{Park et~al.(2020)Park, Harari, Dron, Lester, Thorlund and
  Mills}]{Park2020}
\bibinfo{author}{Park JJ}, \bibinfo{author}{Harari O}, \bibinfo{author}{Dron
  L}, \bibinfo{author}{Lester RT}, \bibinfo{author}{Thorlund K},
  \bibinfo{author}{Mills EJ}.
\newblock \bibinfo{title}{An overview of platform trials with a checklist for
  clinical readers}.
\newblock \emph{\bibinfo{journal}{Journal of Clinical Epidemiology}}
  \bibinfo{year}{2020};.
\bibitem[{Meyer et~al.(2020)Meyer, Mesenbrink, Dunger-Baldauf, Fülle, Glimm,
  Li, Posch and König}]{Meyer2020}
\bibinfo{author}{Meyer EL}, \bibinfo{author}{Mesenbrink P},
  \bibinfo{author}{Dunger-Baldauf C}, \bibinfo{author}{Fülle HJ},
  \bibinfo{author}{Glimm E}, \bibinfo{author}{Li Y}, \bibinfo{author}{Posch M},
  \bibinfo{author}{König F}.
\newblock \bibinfo{title}{The evolution of master protocol clinical trial
  designs: A systematic literature review}.
\newblock \emph{\bibinfo{journal}{Clinical Therapeutics}}
  \bibinfo{year}{2020};\bibinfo{volume}{42}(\bibinfo{number}{7}):\bibinfo{pages}{1330
  -- 1360}.
\newblock DOI:\bibinfo{doi}{https://doi.org/10.1016/j.clinthera.2020.05.010}.
\newblock
  \urlprefix\url{http://www.sciencedirect.com/science/article/pii/S0149291820302447}.
\bibitem[{Israel et~al.(2021)Israel, Denlinger, Bacharier, LaVange, Moore,
  Peters, Georas, Wright, Mauger, Noel et~al.}]{Israel2021}
\bibinfo{author}{Israel E}, \bibinfo{author}{Denlinger LC},
  \bibinfo{author}{Bacharier LB}, \bibinfo{author}{LaVange LM},
  \bibinfo{author}{Moore WC}, \bibinfo{author}{Peters MC},
  \bibinfo{author}{Georas SN}, \bibinfo{author}{Wright RJ},
  \bibinfo{author}{Mauger DT}, \bibinfo{author}{Noel P}, et~al.
\newblock \bibinfo{title}{Precise: Precision medicine in severe asthma: An
  adaptive platform trial with biomarker ascertainment}.
\newblock \emph{\bibinfo{journal}{Journal of Allergy and Clinical Immunology}}
  \bibinfo{year}{2021};\bibinfo{volume}{147}(\bibinfo{number}{5}):\bibinfo{pages}{1594--1601}.
\bibitem[{Saville and Berry(2016)}]{Saville2016}
\bibinfo{author}{Saville BR}, \bibinfo{author}{Berry SM}.
\newblock \bibinfo{title}{{Efficiencies of platform clinical trials: A vision
  of the future}}.
\newblock \emph{\bibinfo{journal}{Clinical Trials}}
  \bibinfo{year}{2016};\bibinfo{volume}{13}(\bibinfo{number}{3}):\bibinfo{pages}{358--366}.
\newblock DOI:\bibinfo{doi}{10.1177/1740774515626362}.
\bibitem[{Brueckner et~al.(2018)Brueckner, Titman, Jaki, Rojek and
  Horby}]{Brueckner2018}
\bibinfo{author}{Brueckner M}, \bibinfo{author}{Titman A},
  \bibinfo{author}{Jaki T}, \bibinfo{author}{Rojek A}, \bibinfo{author}{Horby
  P}.
\newblock \bibinfo{title}{{Performance of different clinical trial designs to
  evaluate treatments during an epidemic}}.
\newblock \emph{\bibinfo{journal}{PLOS ONE}}
  \bibinfo{year}{2018};\bibinfo{volume}{13}(\bibinfo{number}{9}):\bibinfo{pages}{e0203387}.
\newblock DOI:\bibinfo{doi}{10.1371/journal.pone.0203387}.
\bibitem[{Yuan et~al.(2016)Yuan, Guo, Munsell, Lu and Jazaeri}]{Yuan2016}
\bibinfo{author}{Yuan Y}, \bibinfo{author}{Guo B}, \bibinfo{author}{Munsell M},
  \bibinfo{author}{Lu K}, \bibinfo{author}{Jazaeri A}.
\newblock \bibinfo{title}{{MIDAS: a practical Bayesian design for platform
  trials with molecularly targeted agents}}.
\newblock \emph{\bibinfo{journal}{Statistics in Medicine}}
  \bibinfo{year}{2016};\bibinfo{volume}{35}(\bibinfo{number}{22}):\bibinfo{pages}{3892--3906}.
\newblock DOI:\bibinfo{doi}{10.1002/sim.6971}.
\bibitem[{Hobbs et~al.(2018)Hobbs, Chen and Lee}]{Hobbs2018_2}
\bibinfo{author}{Hobbs BP}, \bibinfo{author}{Chen N}, \bibinfo{author}{Lee JJ}.
\newblock \bibinfo{title}{{Controlled multi-arm platform design using
  predictive probability}}.
\newblock \emph{\bibinfo{journal}{Statistical Methods in Medical Research}}
  \bibinfo{year}{2018};\bibinfo{volume}{27}(\bibinfo{number}{1}):\bibinfo{pages}{65--78}.
\newblock DOI:\bibinfo{doi}{10.1177/0962280215620696}.
\bibitem[{Tang et~al.(2018)Tang, Shen and Yuan}]{Tang2018}
\bibinfo{author}{Tang R}, \bibinfo{author}{Shen J}, \bibinfo{author}{Yuan Y}.
\newblock \bibinfo{title}{{ComPAS: A Bayesian drug combination platform trial
  design with adaptive shrinkage}}.
\newblock \emph{\bibinfo{journal}{Statistics in Medicine}}
  \bibinfo{year}{2018};(\bibinfo{number}{March}):\bibinfo{pages}{1--15}.
\newblock DOI:\bibinfo{doi}{10.1002/sim.8026}.
\bibitem[{Ventz et~al.(2017)Ventz, Alexander, Parmigiani, Gelber and
  Trippa}]{Ventz2017}
\bibinfo{author}{Ventz S}, \bibinfo{author}{Alexander BM},
  \bibinfo{author}{Parmigiani G}, \bibinfo{author}{Gelber RD},
  \bibinfo{author}{Trippa L}.
\newblock \bibinfo{title}{{Designing clinical trials that accept new arms: An
  example in metastatic breast cancer}}.
\newblock \emph{\bibinfo{journal}{Journal of Clinical Oncology}}
  \bibinfo{year}{2017};\bibinfo{volume}{35}(\bibinfo{number}{27}):\bibinfo{pages}{3160--3168}.
\newblock DOI:\bibinfo{doi}{10.1200/JCO.2016.70.1169}.
\bibitem[{FDA(2013)}]{FDA_Codevelopment}
\bibinfo{author}{FDA}.
\newblock \bibinfo{title}{{Codevelopment of Two or More New Investigational
  Drugs for Use in Combination. Guidance for Industry (June 2013)}}.
\newblock \bibinfo{year}{2013}.
\newblock \bibinfo{note}{\url{https://www.fda.gov/media/80100/download}
  [Accessed: 2020-06-02]}.
\bibitem[{EMA(2017)}]{EMA_Codevelopment}
\bibinfo{author}{EMA}.
\newblock \bibinfo{title}{{Guideline on clinical development of fixed
  combination medicinal products (March 2017)}}.
\newblock \bibinfo{year}{2017}.
\newblock
  \bibinfo{note}{\url{https://www.ema.europa.eu/en/documents/scientific-guideline/guideline-clinical-development-fixed-combination-medicinal-products-revision-2_en.pdf}
  [Accessed: 2020-06-02]}.
\bibitem[{Jiang et~al.(2020)Jiang, Yan, Thall and Huang}]{Jiang2020}
\bibinfo{author}{Jiang L}, \bibinfo{author}{Yan F}, \bibinfo{author}{Thall PF},
  \bibinfo{author}{Huang X}.
\newblock \bibinfo{title}{Comparing bayesian early stopping boundaries for
  phase ii clinical trials}.
\newblock \emph{\bibinfo{journal}{Pharmaceutical statistics}}
  \bibinfo{year}{2020};\bibinfo{volume}{19}(\bibinfo{number}{6}):\bibinfo{pages}{928--939}.
\bibitem[{Gallo et~al.(2014)Gallo, Mao and Shih}]{Gallo2014}
\bibinfo{author}{Gallo P}, \bibinfo{author}{Mao L}, \bibinfo{author}{Shih VH}.
\newblock \bibinfo{title}{Alternative views on setting clinical trial futility
  criteria}.
\newblock \emph{\bibinfo{journal}{Journal of biopharmaceutical statistics}}
  \bibinfo{year}{2014};\bibinfo{volume}{24}(\bibinfo{number}{5}):\bibinfo{pages}{976--993}.
\bibitem[{Stallard et~al.(2020{\natexlab{b}})Stallard, Todd, Ryan and
  Gates}]{Stallard2020}
\bibinfo{author}{Stallard N}, \bibinfo{author}{Todd S}, \bibinfo{author}{Ryan
  EG}, \bibinfo{author}{Gates S}.
\newblock \bibinfo{title}{Comparison of bayesian and frequentist
  group-sequential clinical trial designs}.
\newblock \emph{\bibinfo{journal}{BMC Medical Research Methodology}}
  \bibinfo{year}{2020}{\natexlab{b}};\bibinfo{volume}{20}(\bibinfo{number}{1}):\bibinfo{pages}{1--14}.
\bibitem[{Dawid(1994)}]{Dawid1994}
\bibinfo{author}{Dawid A}.
\newblock \bibinfo{title}{Selection paradoxes of bayesian inference}.
\newblock \emph{\bibinfo{journal}{Lecture Notes-Monograph Series}}
  \bibinfo{year}{1994};:\bibinfo{pages}{211--220}.
\bibitem[{Senn(2008)}]{Senn2008}
\bibinfo{author}{Senn S}.
\newblock \bibinfo{title}{A note concerning a selection “paradox” of
  dawid's}.
\newblock \emph{\bibinfo{journal}{The American Statistician}}
  \bibinfo{year}{2008};\bibinfo{volume}{62}(\bibinfo{number}{3}):\bibinfo{pages}{206--210}.
\bibitem[{Mandel and Rinott(2009)}]{Mandel2009}
\bibinfo{author}{Mandel M}, \bibinfo{author}{Rinott Y}.
\newblock \bibinfo{title}{A selection bias conflict and frequentist versus
  bayesian viewpoints}.
\newblock \emph{\bibinfo{journal}{The American Statistician}}
  \bibinfo{year}{2009};\bibinfo{volume}{63}(\bibinfo{number}{3}):\bibinfo{pages}{211--217}.
\bibitem[{Viele et~al.(2014)Viele, Berry, Neuenschwander, Amzal, Chen, Enas,
  Hobbs, Ibrahim, Kinnersley, Lindborg et~al.}]{Viele2014}
\bibinfo{author}{Viele K}, \bibinfo{author}{Berry S},
  \bibinfo{author}{Neuenschwander B}, \bibinfo{author}{Amzal B},
  \bibinfo{author}{Chen F}, \bibinfo{author}{Enas N}, \bibinfo{author}{Hobbs
  B}, \bibinfo{author}{Ibrahim JG}, \bibinfo{author}{Kinnersley N},
  \bibinfo{author}{Lindborg S}, et~al.
\newblock \bibinfo{title}{Use of historical control data for assessing
  treatment effects in clinical trials}.
\newblock \emph{\bibinfo{journal}{Pharmaceutical statistics}}
  \bibinfo{year}{2014};\bibinfo{volume}{13}(\bibinfo{number}{1}):\bibinfo{pages}{41--54}.
\bibitem[{Schmidli et~al.(2014)Schmidli, Gsteiger, Roychoudhury, O'Hagan,
  Spiegelhalter and Neuenschwander}]{Schmidli2014}
\bibinfo{author}{Schmidli H}, \bibinfo{author}{Gsteiger S},
  \bibinfo{author}{Roychoudhury S}, \bibinfo{author}{O'Hagan A},
  \bibinfo{author}{Spiegelhalter D}, \bibinfo{author}{Neuenschwander B}.
\newblock \bibinfo{title}{Robust meta-analytic-predictive priors in clinical
  trials with historical control information}.
\newblock \emph{\bibinfo{journal}{Biometrics}}
  \bibinfo{year}{2014};\bibinfo{volume}{70}(\bibinfo{number}{4}):\bibinfo{pages}{1023--1032}.
\bibitem[{Li et~al.(2020)Li, Liu and Snavely}]{Li2020}
\bibinfo{author}{Li W}, \bibinfo{author}{Liu F}, \bibinfo{author}{Snavely D}.
\newblock \bibinfo{title}{Revisit of test-then-pool methods and some practical
  considerations}.
\newblock \emph{\bibinfo{journal}{Pharmaceutical statistics}}
  \bibinfo{year}{2020};\bibinfo{volume}{19}(\bibinfo{number}{5}):\bibinfo{pages}{498--517}.
\bibitem[{Harun et~al.(2020)Harun, Liu and Kim}]{Harun2020}
\bibinfo{author}{Harun N}, \bibinfo{author}{Liu C}, \bibinfo{author}{Kim MO}.
\newblock \bibinfo{title}{Critical appraisal of bayesian dynamic borrowing from
  an imperfectly commensurate historical control}.
\newblock \emph{\bibinfo{journal}{Pharmaceutical statistics}}
  \bibinfo{year}{2020};\bibinfo{volume}{19}(\bibinfo{number}{5}):\bibinfo{pages}{613--625}.
\bibitem[{Fei{\ss}t et~al.(2020)Fei{\ss}t, Krisam and Kieser}]{Feisst2020}
\bibinfo{author}{Fei{\ss}t M}, \bibinfo{author}{Krisam J},
  \bibinfo{author}{Kieser M}.
\newblock \bibinfo{title}{Incorporating historical two-arm data in clinical
  trials with binary outcome: A practical approach}.
\newblock \emph{\bibinfo{journal}{Pharmaceutical statistics}}
  \bibinfo{year}{2020};\bibinfo{volume}{19}(\bibinfo{number}{5}):\bibinfo{pages}{662--678}.
\bibitem[{Bofill~Roig et~al.(2021)Bofill~Roig, Krotka, Glimm, Jacko, K{\"o}nig,
  Magirr, Mesenbrink, Viele and Posch}]{Bofill2021}
\bibinfo{author}{Bofill~Roig M}, \bibinfo{author}{Krotka P},
  \bibinfo{author}{Glimm E}, \bibinfo{author}{Jacko P},
  \bibinfo{author}{K{\"o}nig F}, \bibinfo{author}{Magirr D},
  \bibinfo{author}{Mesenbrink P}, \bibinfo{author}{Viele K},
  \bibinfo{author}{Posch M}.
\newblock \bibinfo{title}{On model-based time trend adjustments in platform
  trials with non-concurrent controls.}
\newblock \emph{\bibinfo{journal}{Working Paper}} \bibinfo{year}{2021};.
\bibitem[{Bretz et~al.(2009)Bretz, Koenig, Brannath, Glimm and
  Posch}]{Bretz2010}
\bibinfo{author}{Bretz F}, \bibinfo{author}{Koenig F},
  \bibinfo{author}{Brannath W}, \bibinfo{author}{Glimm E},
  \bibinfo{author}{Posch M}.
\newblock \bibinfo{title}{Adaptive designs for confirmatory clinical trials}.
\newblock \emph{\bibinfo{journal}{Statistics in medicine}}
  \bibinfo{year}{2009};\bibinfo{volume}{28}(\bibinfo{number}{8}):\bibinfo{pages}{1181--1217}.
\bibitem[{Wason et~al.(2014)Wason, Stecher and Mander}]{Wason2014_2}
\bibinfo{author}{Wason JMS}, \bibinfo{author}{Stecher L},
  \bibinfo{author}{Mander AP}.
\newblock \bibinfo{title}{{Correcting for multiple-testing in multi-arm trials:
  is it necessary and is it done?}}
\newblock \emph{\bibinfo{journal}{Trials}}
  \bibinfo{year}{2014};\bibinfo{volume}{15}(\bibinfo{number}{1}):\bibinfo{pages}{364}.
\newblock DOI:\bibinfo{doi}{10.1186/1745-6215-15-364}.
\bibitem[{Howard et~al.(2018)Howard, Brown, Todd and Gregory}]{Howard2018}
\bibinfo{author}{Howard DR}, \bibinfo{author}{Brown JM}, \bibinfo{author}{Todd
  S}, \bibinfo{author}{Gregory WM}.
\newblock \bibinfo{title}{{Recommendations on multiple testing adjustment in
  multi-arm trials with a shared control group}}.
\newblock \emph{\bibinfo{journal}{Statistical Methods in Medical Research}}
  \bibinfo{year}{2018};\bibinfo{volume}{27}(\bibinfo{number}{5}):\bibinfo{pages}{1513--1530}.
\newblock DOI:\bibinfo{doi}{10.1177/0962280216664759}.
\bibitem[{Stallard et~al.(2019)Stallard, Todd, Parashar, Kimani and
  Renfro}]{Stallard2019}
\bibinfo{author}{Stallard N}, \bibinfo{author}{Todd S},
  \bibinfo{author}{Parashar D}, \bibinfo{author}{Kimani PK},
  \bibinfo{author}{Renfro LA}.
\newblock \bibinfo{title}{{On the need to adjust for multiplicity in
  confirmatory clinical trials with master protocols}}.
\newblock \emph{\bibinfo{journal}{Annals of Oncology}}
  \bibinfo{year}{2019};\bibinfo{volume}{30}(\bibinfo{number}{4}):\bibinfo{pages}{506--509}.
\newblock DOI:\bibinfo{doi}{10.1093/annonc/mdz038}.
\bibitem[{Collignon et~al.(2020)Collignon, Gartner, Haidich, James~Hemmings,
  Hofner, P{\'e}tavy, Posch, Rantell, Roes and Schiel}]{Collignon2020}
\bibinfo{author}{Collignon O}, \bibinfo{author}{Gartner C},
  \bibinfo{author}{Haidich AB}, \bibinfo{author}{James~Hemmings R},
  \bibinfo{author}{Hofner B}, \bibinfo{author}{P{\'e}tavy F},
  \bibinfo{author}{Posch M}, \bibinfo{author}{Rantell K}, \bibinfo{author}{Roes
  K}, \bibinfo{author}{Schiel A}.
\newblock \bibinfo{title}{Current statistical considerations and regulatory
  perspectives on the planning of confirmatory basket, umbrella, and platform
  trials}.
\newblock \emph{\bibinfo{journal}{Clinical Pharmacology \& Therapeutics}}
  \bibinfo{year}{2020};\bibinfo{volume}{107}(\bibinfo{number}{5}):\bibinfo{pages}{1059--1067}.
\bibitem[{Berry(2020)}]{berry2020potential}
\bibinfo{author}{Berry SM}.
\newblock \bibinfo{title}{Potential statistical issues between designers and
  regulators in confirmatory basket, umbrella, and platform trials}.
\newblock \emph{\bibinfo{journal}{Clinical Pharmacology \& Therapeutics}}
  \bibinfo{year}{2020};\bibinfo{volume}{108}(\bibinfo{number}{3}):\bibinfo{pages}{444--446}.
\bibitem[{Meyer and Kumaus(2021)}]{Shinyserver}
\bibinfo{author}{Meyer EL}, \bibinfo{author}{Kumaus C}.
\newblock \bibinfo{title}{Shiny app for simulation result exploration and
  visualization}.
\newblock \bibinfo{year}{2021}.
\newblock
  \bibinfo{note}{\url{https://sny.cemsiis.meduniwien.ac.at/~zrx5rdf/oWer32/}
  [Accessed: 2021-10-12]}.
\bibitem[{Meyer et~al.(2021)Meyer, Mesenbrink, Mielke, Parke, Evans and
  K{\"o}nig}]{Meyer2021}
\bibinfo{author}{Meyer EL}, \bibinfo{author}{Mesenbrink P},
  \bibinfo{author}{Mielke T}, \bibinfo{author}{Parke T}, \bibinfo{author}{Evans
  D}, \bibinfo{author}{K{\"o}nig F}.
\newblock \bibinfo{title}{Systematic review of available software for multi-arm
  multi-stage and platform clinical trial design}.
\newblock \emph{\bibinfo{journal}{Trials}}
  \bibinfo{year}{2021};\bibinfo{volume}{22}(\bibinfo{number}{1}):\bibinfo{pages}{1--14}.
\bibitem[{Bai et~al.(2020)Bai, Deng and Liu}]{bai2020multiplicity}
\bibinfo{author}{Bai X}, \bibinfo{author}{Deng Q}, \bibinfo{author}{Liu D}.
\newblock \bibinfo{title}{Multiplicity issues for platform trials with a shared
  control arm}.
\newblock \emph{\bibinfo{journal}{Journal of Biopharmaceutical Statistics}}
  \bibinfo{year}{2020};:\bibinfo{pages}{1--13}.
\bibitem[{Parker and Weir(2020)}]{parker2020non}
\bibinfo{author}{Parker RA}, \bibinfo{author}{Weir CJ}.
\newblock \bibinfo{title}{Non-adjustment for multiple testing in multi-arm
  trials of distinct treatments: Rationale and justification}.
\newblock \emph{\bibinfo{journal}{Clinical Trials}}
  \bibinfo{year}{2020};\bibinfo{volume}{17}(\bibinfo{number}{5}):\bibinfo{pages}{562--566}.
\bibitem[{Bretz and Koenig(2020)}]{bretz2020commentary}
\bibinfo{author}{Bretz F}, \bibinfo{author}{Koenig F}.
\newblock \bibinfo{title}{Commentary on parker and weir}.
\newblock \emph{\bibinfo{journal}{Clinical Trials}}
  \bibinfo{year}{2020};\bibinfo{volume}{17}(\bibinfo{number}{5}):\bibinfo{pages}{567--569}.
\bibitem[{Posch and K{\"o}nig(2020)}]{posch2020commentare}
\bibinfo{author}{Posch M}, \bibinfo{author}{K{\"o}nig F}.
\newblock \bibinfo{title}{Are p-values useful to judge the evidence against the
  null hypotheses in complex clinical trials? a comment on “the role of
  p-values in judging the strength of evidence and realistic replication
  expectations”}.
\newblock \emph{\bibinfo{journal}{Statistics in Biopharmaceutical Research}}
  \bibinfo{year}{2020};:\bibinfo{pages}{1--3}.
\bibitem[{Wason and Robertson(2021)}]{Wason2020}
\bibinfo{author}{Wason JM}, \bibinfo{author}{Robertson DS}.
\newblock \bibinfo{title}{Controlling type i error rates in multi-arm clinical
  trials: A case for the false discovery rate}.
\newblock \emph{\bibinfo{journal}{Pharmaceutical Statistics}}
  \bibinfo{year}{2021};\bibinfo{volume}{20}(\bibinfo{number}{1}):\bibinfo{pages}{109--116}.
\bibitem[{Viele et~al.(2020)Viele, Saville, McGlothlin and Broglio}]{Viele2020}
\bibinfo{author}{Viele K}, \bibinfo{author}{Saville BR},
  \bibinfo{author}{McGlothlin A}, \bibinfo{author}{Broglio K}.
\newblock \bibinfo{title}{Comparison of response adaptive randomization
  features in multiarm clinical trials with control}.
\newblock \emph{\bibinfo{journal}{Pharmaceutical statistics}}
  \bibinfo{year}{2020};\bibinfo{volume}{19}(\bibinfo{number}{5}):\bibinfo{pages}{602--612}.
\bibitem[{Kammer(2020)}]{Kammer2020}
\bibinfo{author}{Kammer M}.
\newblock \bibinfo{title}{looplot: A package for creating nested loop plots}.
\newblock
  \bibinfo{howpublished}{\url{https://github.com/matherealize/looplot}};
  \bibinfo{year}{2020}.
\bibitem[{R{\"u}cker and Schwarzer(2014)}]{Ruecker2014}
\bibinfo{author}{R{\"u}cker G}, \bibinfo{author}{Schwarzer G}.
\newblock \bibinfo{title}{Presenting simulation results in a nested loop plot}.
\newblock \emph{\bibinfo{journal}{BMC medical research methodology}}
  \bibinfo{year}{2014};\bibinfo{volume}{14}(\bibinfo{number}{1}):\bibinfo{pages}{129}.

\end{thebibliography}
\clearpage


\normalsize
\appendix

\section{Dynamic Borrowing} \label{dynamic_borrowing}

\begin{sidewaysfigure}[ht]
\includegraphics[scale=0.45]{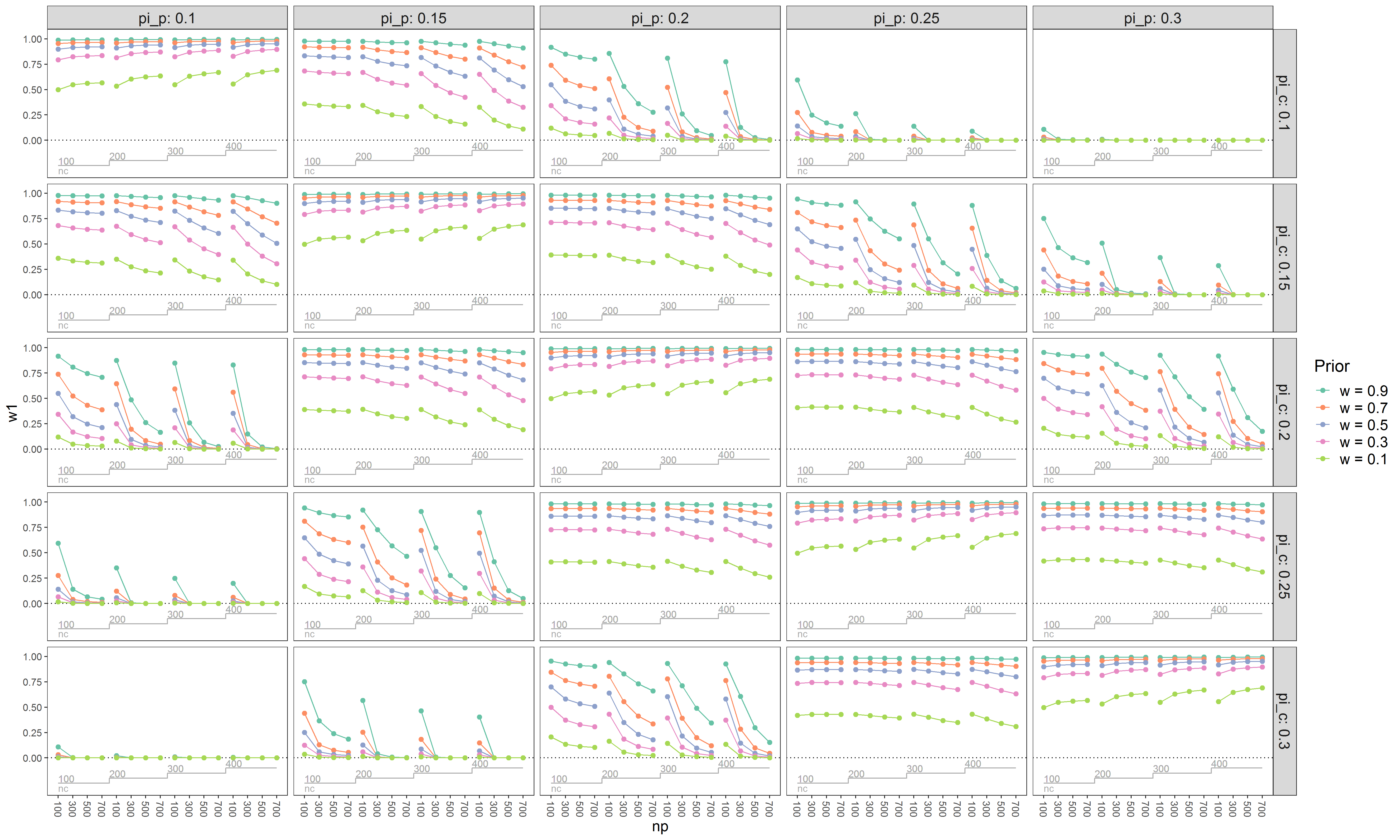}
\caption{Impact of the sample sizes and observed response rates in the cumulative observed cohort and cumulative pooled observed data from other cohorts in the platform trial ($n_c$, $\pi_c = \frac{k_c}{n_c}$ and $n_p$, $\pi_p = \frac{k_c^p}{n_c^p}$ respectively), as well as the prior $w$ on the weight $w_1$. Please note in the figure the label ``pi\_c" is used for $\pi_c$, ``pi\_p" is used for $\pi_p$, ``nc" is used for $n_c$ and ``np" is used for $n_p$. When the observed response rates in the two groups are the same, the amount of data sharing increases with $w$ and increasing sample size in both groups. When the observed response rates are different, the amount of data sharing decreases with decreasing $w$ and increasing sample size in both groups. When the observed response rates are very different, the amount of data sharing is nearly 0, even when the sample sizes are small and the prior $w$ was chosen to be 0.9. When e.g. the prior $w$ is set to 0.1, even in case of equality of $\pi_c$ and $\pi_p$, the actual borrowing $w_1$ does not increase beyond approximately 0.7 for the chosen sample sizes. For all results, the Beta prior parameters were set to 0.5. Visualization is based on the \textbf{looplot} package \citep{Kammer2020}, which implements the visualisation presented by \citet{Ruecker2014}. Please note that e.g. in the plot in the bottom left corner, most lines are overlapping.}
\label{fig:dyn_bor}
\end{sidewaysfigure}

Assume at any given time during the course of the platform trial we want to compute the posterior probability for either the backbone monoterhapy or SoC response rate $\theta_c$ in cohort $c$ using a ``dynamic borrowing" approach, in particular using a robust mixture prior (see \citet{Schmidli2014}), in which we both use the cumulative observed data of cohort $c$ (denoted by $(n_c, k_c)$), as well as the cumulative \textbf{p}ooled observed data from other cohorts in the platform trial (denoted by $(n_c^p, k_c^p)$), where $n$ denotes the sample size and $k$ the number of responders. The data $(n_c^p, k_c^p)$ might include pooled data from both earlier cohorts of the same platform trial, or concurrent data from concurrent cohorts of the same platform trial, but does not include data that has not yet been observed. E.g. when looking at figure \ref{fig:sharing}, $(n_c, k_c)$ is the data from cohort 3 and $(n_c^p, k_c^p)$ is the pooled data from cohorts 1 and 2 (for the backbone monotherapy and SoC arms respectively). Ultimately our goal is to have a posterior for $\theta_c$ that consists of two weighted Beta distributions, one informed by both $(n_c, k_c)$ and $(n_c^p, k_c^p)$ with weight $w_1$ and the other informed only by $(n_c, k_c)$ with weight $w_2$ ($w_1 + w_2 = 1$), in other terms
\begin{align}
\label{posterior}
\pi(\theta_c) = w_1*f(y, k_c^p + k_c + \alpha_c^p, n_c^p + n_c - k_c^p - k_c + \beta_c^p) + w_2*f(y, k_c + \alpha_c, n_c - k_c + \beta_c),
\end{align}
where $f(\cdot, \alpha, \beta)$ denotes the probability density function of a Beta distribution with parameters $\alpha$ and $\beta$ and $ \pi(\theta_c)$ denotes the density of the posterior distribution of $\theta_c$. The weights then represent the degree of ``borrowing''. Furthermore, the a priori unknown degree of borrowing (i.e. $w_1$ and $w_2$) should be based on $(n_c, k_c)$ and $(n_c^p, k_c^p)$. \newline
Our robust mixture prior for $\theta_c$ conditional on this data is given by
\begin{align}
\label{prior}
\pi^{RMP}(\theta_c|n_c^p,k_c^p) &= w*f(\theta_c,k_c^p + \alpha_c, n_c^p-k_c^p+\beta_c) + (1-w)*f(\theta_c, \alpha_c, \beta_c) =& \nonumber \\ \nonumber \\ &= w \frac{\theta_c^{k_c^p+\alpha_c-1}*(1-\theta_c)^{n_c^p-k_c^p+\beta_c-1}}{B(k_c^p+\alpha_c, n_c^p - k_c^p + \beta_c)} + (1-w)\frac{\theta_c^{\alpha_c-1}*(1-\theta_c)^{\beta_c-1}}{B(\alpha_c, \beta_c)}, &
\end{align}
where $B(\alpha, \beta)$ denotes the Beta function at $(\alpha,\beta)$ and $\alpha_c$ and $\beta_c$ are the parameters reflecting the current information about $\theta_c$ (both set to $0.5$ through our simulations). Please note that this posterior is not iteratively derived, but rather "rebuilt" from scratch every time an analysis is conducted. Please further note that $w_1$ and $w_2$ (equation \ref{posterior}) depend on $w$ (equation \ref{prior}) and $w$ needs to be chosen a priori. The posterior distribution of $\theta_c$ is now derived as:
\begin{align}
\pi(\theta_c) &\propto \pi^{RMP}(\theta_c|n_c^p, k_c^p)*\theta_c^{k_c}*(1-\theta_c)^{n_c-k_c}  =& \nonumber \\ \nonumber \\ &= w \frac{\theta_c^{k_c^p+k_c + \alpha_c-1}*(1-\theta_c)^{n_c^p + n_c-k_c^p-k_c+\beta_c-1}}{B(k_c^p +\alpha_c, n_c^p - k_c^p + \beta_c)} + (1-w) \frac{\theta_c^{k_c + \alpha_c-1}*(1-\theta_c)^{n_c - k_c +\beta_c-1}}{B(\alpha_c, \beta_c)}. &
\end{align}
In order for this to be a distribution, we need to make sure it integrates to 1. An option which captures the idea of dynamic borrowing in the sense that the more similar the posterior of $\theta_c$ based on $(n_c, k_c)$ is to the posterior of $\theta_c$ based on $(n_c^p, k_c^p)$, the more borrowing should be done - is to set
\begin{align}
w_1 = \frac{w\frac{B(k_c^p+k_c+\alpha_c, n_c^p+n_c-k_c^p-k_c+\beta_c)}{B(k_c^p+\alpha_c, n_c^p-k_c^p+\beta_c)}}{w\frac{B(k_c^p+k_c+\alpha_c, n_c^p+n_c-k_c^p-k_c+\beta_c)}{B(k_c^p+\alpha_c, n_c^p-k_c^p+\beta_c)} + (1-w)\frac{B(k_c+\alpha_c, n_c-k_c+\beta_c)}{B(\alpha_c, \beta_c)}}
\end{align}
and
\begin{align}
w_2 = \frac{(1-w)\frac{B(k_c+\alpha_c, n_c-k_c+\beta_c)}{B(\alpha_c, \beta_c)}}{w\frac{B(k_c^p+k_c+\alpha_c, n_c^p+n_c-k_c^p-k_c+\beta_c)}{B(k_c^p+\alpha_c, n_c^p-k_c^p+\beta_c)} + (1-w)\frac{B(k_c+\alpha_c, n_c-k_c+\beta_c)}{B(\alpha_c, \beta_c)}} .
\end{align}
This corresponds to the case in which we divide the unscaled posterior in (3) by the factor
\begin{align}
w\frac{B(k_c^p+k_c+\alpha_c, n_c^p+n_c-k_c^p-k_c+\beta_c)}{B(k_c^p+\alpha_c, n_c^p-k_c^p+\beta_c)} + (1-w)\frac{B(k_c+\alpha_c, n_c-k_c+\beta_c)}{B(\alpha_c, \beta_c)}.
\end{align}
The posterior distribution is now given by
\begin{align}
\pi(\theta_c) = \frac{w\frac{\theta_c^{k_c^p+k_c + \alpha_c-1}*(1-\theta_c)^{n_c^p + n_c-k_c^p-k_c+\beta_c-1}}{B(k_c^p +\alpha_c, n_c^p - k_c^p + \beta_c)} + (1-w)\frac{\theta_c^{k_c + \alpha_c-1}*(1-\theta_c)^{n_c - k_c +\beta_c-1}}{B(\alpha_c, \beta_c)}}{w\frac{B(k_c^p+k_c+\alpha_c, n_c^p+n_c-k_c^p-k_c+\beta_c)}{B(k_c^p+\alpha_c, n_c^p-k_c^p+\beta_c)} + (1-w)\frac{B(k_c+\alpha_c, n_c-k_c+\beta_c)}{B(\alpha_c, \beta_c)}} .
\end{align}
Since $w_1 + w_2 = 1$, we can be sure that this yields a proper distribution. Choosing the weights $w_1$ and $w_2$ in this way now leads to the borrowing being dynamic. In figure \ref{fig:dyn_bor} we investigate the impact of the sample sizes and observed response rates in the cumulative observed cohort and cumulative pooled observed data from other cohorts in the platform trial, as well as the prior $w$ on the weight $w_1$.


Finally, the parameters of the Beta posterior are derived as

\begin{align}
\alpha_{eff} = w_{1}*(k_{c}^p + k_{c} + \alpha_{c}) + w_{2}*(k_{c} + \alpha_{c})
\end{align}

\begin{align}
\beta_{eff} = w_{1}*((n_c + n_c^p) - (k_{c}^p + k_{c}) + \alpha_{c}) + w_{2}*(n_c - k_{c} + \alpha_{c})
\end{align}

Despite not being investigated in this manuscript, the simulation software also features frequentist decision rules based on 2x2 contingency tables and therefore expects the effective sample size and number of responders to be integers. We therefore simplify and assume the effective sample size to be $\lfloor \alpha_{eff} + \beta_{eff} \rceil$ and the effective number of responders to be $\lfloor \alpha_{eff} \rceil$. The whole procedure is applied for backbone monotherapy and SoC separately.

\clearpage


\section{Different target product profiles} \label{zeta}

\begin{figure}[ht]
\includegraphics[scale=0.70]{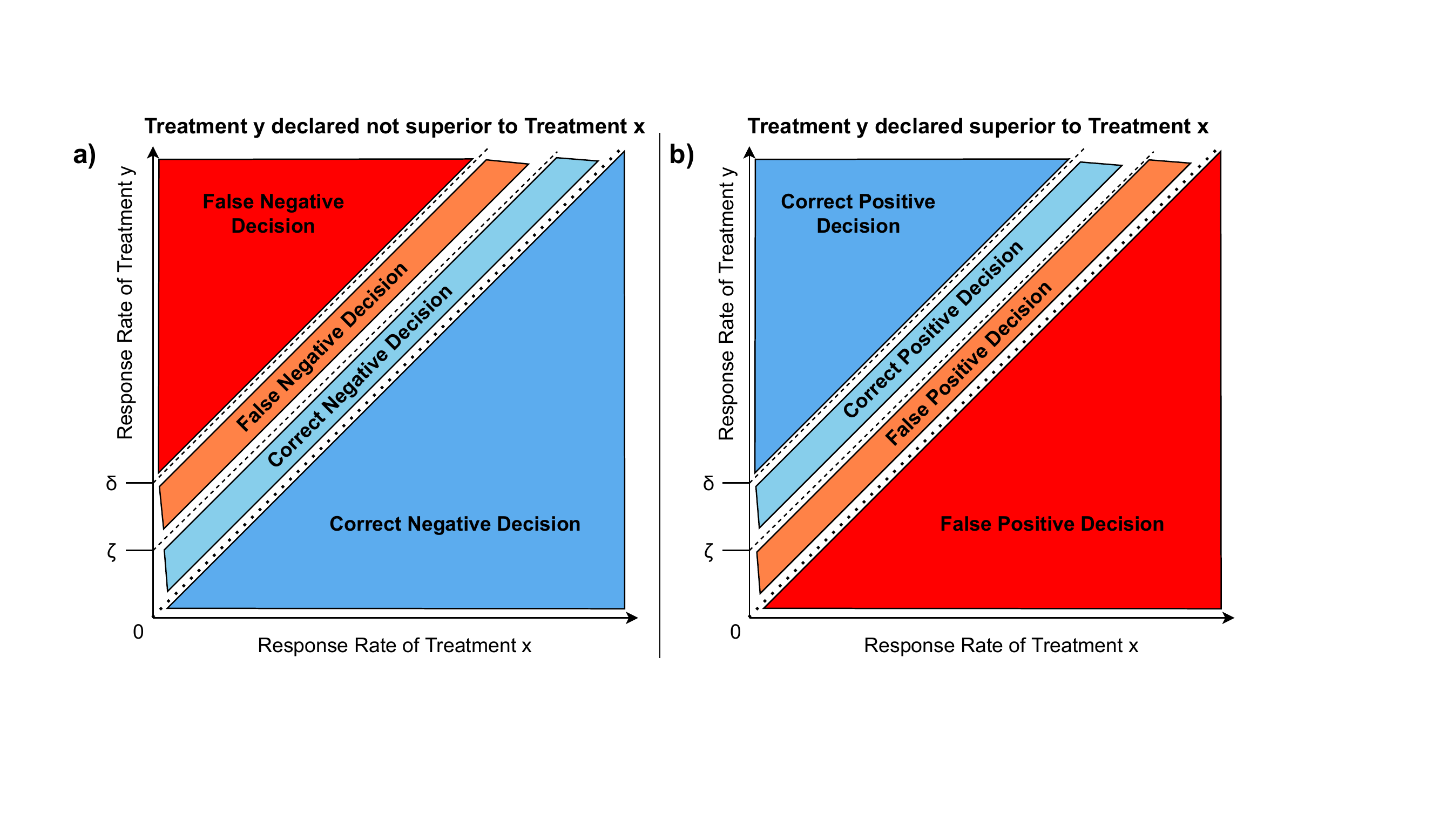}
\caption{After a pair-wise comparison of treatment x against treatment y, whereby the decision rules required the response rate of treatment y ($\pi_y$) to be superior to the response rate of treatment x ($\pi_x$) by a margin of $\delta$ (i.e. $\pi_y \geq \pi_x + \delta$), treatment y was either declared superior (subfigure b) or not superior (subfigure a) to treatment x. A target product profile was defined for treatment y, whereby the aim was for the response rate of treatment y to be superior to the response rate of treatment x by a margin of $\zeta$ (i.e. $\pi_y \geq \pi_x + \zeta$). For simplicity, we assume $\delta \geq \zeta$, although this approach also works when $\delta < \zeta$. When in truth $\pi_y < \pi_x + \zeta$, the decision is either a correct negative decision (subfigure a) or a false positive decision (subfigure b). When in truth $\pi_y \geq \pi_x + \zeta$, the decision is either a false negative decision (subfigure a) or a correct positive decision (subfigure b). Please note that in the usual defintion of a type 1 error, $\zeta = 0$.}
\label{fig:tpp}
\end{figure}

All considered pair-wise decision rules are of the form ``GO", if $P(\pi_y > \pi_x + \delta| Data) > \gamma$ (see section \ref{Decision Rules}). In section \ref{OCs}, we assumed that the target product profile would always require the response rate of drug y to be greater than the response rate of drug x by any margin to consider the alternative hypothesis to hold. This can be extended to allow the target product profile to specify any required margin $\zeta$, such that if and only if the response rate of drug y is truly better than the response rate of drug x by a margin of at least $\zeta$ do we consider the alternative hypothesis to hold. For every pair-wise comparison, this margin $\zeta$ can, but does not have to, coincide with the $\delta$ used. This is visualized for a single pair-wise comparison in Figure \ref{fig:tpp}. As a reminder, for all of the pair-wise comparisons in our simulations, we set $\zeta = 0$.  In the \textbf{CohortPlat} package, a different margin $\zeta$ can be chosen both for the comparison of combination against the monotherapies and the comparison of the monotherapies against SoC. 

\clearpage

\section{Treatment Efficacy Scenarios} \label{Efficacy}

In the main text, we showed nearly exclusively selected results of one set of assumptions regarding the treatment effect of the monotherapies and the combination treatment (setting 1). In general, we only investigated treatment effect assumptions based on risk-ratios, whereby we randomly and separately draw the risk-ratio for each of the monotherapies with respect to the SoC treatment. For the combination treatment, we randomly draw from a range of interaction effects, which could result in additive, synergistic or antagonistic effects of a specified magnitude. Some scenarios might be more realistic for a given drug development program than others, however we felt that the broad range of scenarios will allow to investigate the impact and interaction of the various simulation parameters and assumptions on the operating characteristics. Let $\pi_x$ denote the probability of a patient on therapy $x$ to have a successful treatment outcome (binary), i.e. the response-rate, and $T_x$ denote a discrete random variable. \newline

In detail, every time a new cohort enters the platform, we firstly fix the SoC response-rate:

$$
\begin{array}{l}
\pi_{SoC} \in [0, 1]
 \end{array}
$$

Then we assign the treatment effect in terms of risk-ratios for the backbone monotherapy (monotherapy A), which is the same across all cohorts:

$$
\begin{array}{l}
\pi_{MonoA} = \pi_{SoC} * \gamma_{MonoA}, \gamma_{MonoA} \sim T_{MonoA}
 \end{array}
$$

Then we randomly draw the treatment effect in terms of risk-ratios for the add-on monotherapy (monotherapy B):

$$
\begin{array}{l}
\pi_{MonoB} = \pi_{SoC} * \gamma_{MonoB}, \gamma_{MonoB} \sim T_{MonoB}
 \end{array}
$$

Finally, after knowing the treatment effects of both monotherapies, we randomly drew an interaction effect for the combination treatment:

$$
\begin{array}{l}
\pi_{Combo} = \pi_{SoC} * (\gamma_{MonoA} * \gamma_{MonoB})*\gamma_{Combo} , \gamma_{Combo} \sim T_{Combo}
 \end{array}
$$

Depending on the scenario, the distribution functions can have all the probability mass on one value, i.e. the assignment of treatment effects and risk-ratios is not necessarily random. Please further note that while the treatment effects were specified in terms of risks and risk-ratios, the Bayesian decision rules were specified in terms of response rates. Two settings characterize global null hypotheses, six settings characterize an efficacious backbone monotherapy with varying degrees of add-on mono and combination therapy efficacy, two settings characterize an efficacious backbone with varying degrees of random add-on mono and combination therapy efficacy, two settings characterize either the global null hypothesis or efficacious mono and combination therapies, but with an underlying time-trend, and two settings were run as sensitivity analyses with increased standard-of-case response rates. The different treatment efficacy settings are summarized in table \ref{tab:settings}. 

\begin{sidewaystable}
\footnotesize
  \caption{Overview of different treatment effect assumptions. The priors $T_{MonoA}$, $T_{MonoB}$ and $T_{Comb}$ for $\gamma_{MonoA}$, $\gamma_{MonoB}$ and $\gamma_{Comb}$ as described in section \ref{Efficacy} are all pointwise with a support of 1,2 or 3 different points (each with probability ``p'') and result in effective response rates $\pi_{SoC}$, $\pi_{MonoA}$, $\pi_{MonoB}$ and $\pi_{Comb}$. Only results of setting 1 are shown in the main text, the rest in the supplements.}
  \label{tab:settings}
\begin{tabularx}{\textwidth}{l|c|c|c|c|X}
\hline
Setting & $\pi_{SoC}$ & \makecell{$\pi_{MonoA}$ \\ ($\gamma_{MonoA}$)} & \makecell{$\pi_{MonoB}$ \\ ($\gamma_{MonoB}$)} & \makecell{$\pi_{Combo}$ \\ ($\gamma_{Combo}$)} & Description \\
\hline

1       &       0.10        &       0.20 (2)       &      \makecell{ 
0.10 (1) with p 0.5 \\
0.20 (2) with p 0.5
}     &         \makecell{ 
0.20 (1) if $\gamma_{MonoB} = 1$ \\
0.40 (1) if $\gamma_{MonoB} = 2$
}        &   backbone monotherapy superior to SoC, add-on monotherapy has 50:50 chance to be superior to SoC; in case add-on monotherapy not superior to SoC, combination therapy as effective as backbone monotherapy, otherwise combination therapy significantly better than monotherapies  \\
\hline

2  &  0.10 & 0.20 (2)  & 0.10 (1)  &  0.20 (1)  &  backbone monotherapy superior to SoC, but add-on monotherapy not superior to SoC and combination therapy not better than backbone monotherapy    \\
\hline

3  &  0.10  & 0.20 (2) &  0.10 (1) &  0.30 (1.5) &  backbone monotherapy superior to SoC and combination therapy superior to backbone monotherapy, but add-on monotherapy not superior to SoC   \\
\hline

4  & 0.10  &  0.20 (2)  &  0.10 (1)  & 0.40 (2)  &  backbone monotherapy superior to SoC and combination therapy superior to backbone monotherapy (increased combination treatment effect compared to setting 4), but add-on monotherapy not superior to SoC   \\
\hline

5  &   0.10  &  0.20 (2) & 0.20 (2)  &  0.20 (0.5)  &  both monotherapies are superior to SoC, but combination therapy is not better than monotherapies   \\
\hline

6  &  0.10  & 0.20 (2)  & 0.20 (2)   & 0.30 (0.75)  &  both monotherapies are superior to SoC and combination therapy is better than monotherapies   \\
\hline

7  & 0.10  &   0.20 (2) & 0.20 (2) & 0.40 (1) &  both monotherapies are superior to SoC and combination therapy is superior to monotherapies (increased combination treatment effect compared to setting 7)   \\
\hline

8  &  0.10  &  0.10 (1)   & 0.10 (1)  &  0.10 (1)  &    global null hypothesis         \\
\hline

9  &  0.20  &  0.20 (1)  & 0.20 (1)  & 0.20 (1) &  global null hypothesis with higher response rates       \\
\hline

10       &       0.10        &       0.20 (2)       &      \makecell{ 
0.10 (1) with p 0.5 \\
0.20 (2) with p 0.5
}     &         \makecell{ 
0.20*$\gamma_{MonoB}$*0.5 (0.5) with p $\frac{1}{3}$ \\
0.20*$\gamma_{MonoB}$*1 (1) with p $\frac{1}{3}$ \\
0.20*$\gamma_{MonoB}$*1.5 (1.5) with p $\frac{1}{3}$
}        &   backbone monotherapy superior to SoC, add-on monotherapy has 50:50 chance to be superior to SoC; combination therapy interaction effect can either be antagonistic/non-existent, additive or synergistic (with equal probabilities) \\
\hline

11       &   \makecell{0.10 + \\ 0.03*(c-1) }        &       \makecell{0.10 + \\ 0.03*(c-1) (1) }       &  0.10 + 0.03*(c-1) (1)     &        0.10 + 0.03*(c-1) (1)   &   time-trend null scenario; every new cohort (first cohort $c=1$, second cohort $c=2$, ...) will have SoC response rate that is by 3\%-points higher than that of the previous cohort  \\
\hline

12      &       \makecell{0.10 + \\ 0.03*(c-1) }      &    \makecell{0.20 + \\ 0.03*(c-1) (2) }     &  0.20 + 0.03*(c-1) (2)     &        0.40 + 0.03*(c-1) (1)   &   time-trend scenario, whereby monotherapies superior to SoC and combination therapy superior to monotherapies; every new cohort (first cohort $c=1$, second cohort $c=2$, ...) will have SoC response rate that is by 3\%-points higher than that of the previous cohort  \\
\hline

13  &   0.20  &  0.30 (1.5) & 0.30 (1.5)  &  0.40 ($\frac{8}{9}$)  &  analogous to setting 7, but SoC response rate is 20\%   \\
\hline

14  &   0.20  &  0.30 (1.5) & 0.30 (1.5)  &  0.50 ($\frac{10}{9}$)  &  analogous to setting 8, but SoC response rate is 20\%    \\
\hline

\end{tabularx}
\end{sidewaystable}

\clearpage

\end{document}